\RequirePackage[left, pagewise]{lineno}
\documentclass[floatfix,aps,prd,twocolumn,letterpaper,lengthcheck,superscriptaddress,amssymb,amsmath,amsfonts,nofootinbib,altaffilletter]{revtex4}
\usepackage{color}
\usepackage[dvipsnames]{xcolor}
\usepackage{calc}
\usepackage{mathtools,graphicx}
\usepackage{braket}
\usepackage{pifont}
\usepackage{bm}
\usepackage{microtype}
\usepackage{booktabs}
\usepackage{times}
\usepackage[varg]{txfonts}
\usepackage[colorlinks, pdfborder={0 0 0}]{hyperref}
\usepackage[utf8]{inputenc}
\usepackage{paralist}
\usepackage[normalem]{ulem}
\usepackage{multirow}
\usepackage{etoolbox}
\usepackage{xstring}
\usepackage{xparse}
\usepackage{dcolumn}
\usepackage{scrextend}
\usepackage{textgreek}
\usepackage{makerobust}
\usepackage{caption,subcaption}
\definecolor{LinkColor}{rgb}{0.75, 0, 0}
\definecolor{CiteColor}{rgb}{0, 0.5, 0.5}
\definecolor{UrlColor}{rgb}{0, 0, 0.75}
\hypersetup{linkcolor=LinkColor}
\hypersetup{citecolor=CiteColor}
\hypersetup{urlcolor=UrlColor}
\allowdisplaybreaks
\DeclareFontFamily{OT1}{pzc}{}
\DeclareFontShape{OT1}{pzc}{m}{it}{<-> s * [1.10] pzcmi7t}{}
\DeclareMathAlphabet{\mathpzc}{OT1}{pzc}{m}{it}

\graphicspath{{./figures/}}

\begin{document}
\title{Impact of noise transients on low latency gravitational-wave event localisation}
\author{Ronaldas Macas}
\affiliation{University of Portsmouth, Portsmouth, PO1 3FX, UK}
\author{Joshua Pooley}
\affiliation{University of Portsmouth, Portsmouth, PO1 3FX, UK}
\author{Laura K. Nuttall}
\affiliation{University of Portsmouth, Portsmouth, PO1 3FX, UK}
\author{Derek Davis}
\affiliation{LIGO, California Institute of Technology, Pasadena, CA 91125, USA}
\author{Martin J. Dyer}
\affiliation{The University of Sheffield, Sheffield S3 7RH, UK}
\author{Yannick Lecoeuche}
\affiliation{University of British Columbia, Vancouver, BC V6T 1Z4, Canada}
\author{Joseph D. Lyman}
\affiliation{University of Warwick, Gibbet Hill Road, Coventry CV4 7AL, UK}
\author{Jess McIver}
\affiliation{University of British Columbia, Vancouver, BC V6T 1Z4, Canada}
\author{Katherine Rink}
\affiliation{University of Massachusetts, Dartmouth, Massachusetts 02747, USA}

\begin{abstract}
    Gravitational-wave (GW) data contains non-Gaussian noise transients called ‘glitches’. 
During the third LIGO-Virgo observing run about 24\% of all gravitational-wave candidates were in the vicinity of a glitch, while even more events could be affected in future observing runs due to increasing detector sensitivity.
This poses a problem since glitches can affect the estimation of GW source parameters, including sky localisation, which is crucial to identify an electromagnetic counterpart.
This is first of a kind study that evaluates the importance of relative glitch positioning in time with respect to a GW signal.
In this paper we estimate how much sky localisation is affected by a nearby glitch in low latency.
We injected binary black hole (BBH), binary neutron star (BNS) and neutron star-black hole (NSBH) signals into data containing three different classes of glitches: blips, thunderstorms and fast scatterings.
The impact of these glitches was assessed by estimating the number of tile pointings that a telescope would need to search over until the true sky location of an event is observed. 
We find that blip glitches generally do not affect the localisation of our tested GW signals, however in very rare cases of a blip glitch overlap with a BBH or a NSBH signal can cause the true position of the event to lie well outside the 90\% computed sky-localisation, severely compromising electromagnetic follow-up.
Thunderstorm glitches have a noticeable impact on BBH and NSBH events, especially if there is no third interferometer.
In such cases we find that the EM follow-up efforts with telescopes as large as $20\,\text{deg}^2$ field-of-view (FOV) are affected.
Observing BBH and NBSH signals with three-detector network reduces the bias in sky localisation caused by thunderstorm glitches, making the bias to affect only small (FOV=$1\,\text{deg}^2$) telescopes.
BNS events appear to be not affected by thunderstorm glitches.
Fast scattering glitches have no impact on the low latency localisation of BBH and BNS signals. 
For NSBH signals observed with two-detector network, the sky localisation bias due to fast scattering glitches is significant enough to affect even large (FOV=$20\,\text{deg}^2$) telescopes.
Observing NSBH signals with three interferometers reduces the bias such that it impacts only small (FOV=$1\,\text{deg}^2$) telescopes.
\end{abstract}

\maketitle

\section{Introduction}
\label{intro}

Over the last six years LIGO and Virgo have detected 90 gravitational-wave (GW) candidates~\cite{o3b_cat}, all from the merger of neutron stars and black holes.
However, a number of these candidates coincided with transient noise from gravitational-wave detectors.
Out of 74 GW candidates detected in the third LIGO-Virgo observing run (O3), 18 of them ($24\%$) had nearby non-Gaussian noise artifacts in one or more detectors~\citep{o3a_cat, o3b_cat}.
This transient noise did not impact the detection of these candidate events but had to be mitigated before the source parameters could be estimated.

In an ideal case, gravitational-wave interferometer data can be described as stationary (\textit{i.e.}~does not vary in time) and Gaussian.
However in reality the data is neither.
It often contains periods of non-stationarity as well as non-Gaussian noise transients or `glitches' \citep{dc_general, blip_rate}.
Both types of noise can affect analyses which estimate source parameters \citep{glitch_pe_3, glitch_pe_4, glitch_pe_5} and certain types of glitches can be mistaken for true GW signals \citep{glitch_pe_1, glitch_pe_2, glitch_pe_3, blips}, yet there has been no work that assesses the impact of glitches overlapping GW signals.
There are many examples which create such noise in the detectors, ranging from natural ones, such as earthquakes~\citep{eq_1, eq_2}, to human-made ones, like a helicopter flying over an interferometer~\cite{helicopter}.

The most well known example of a GW event affected by transient noise is binary neutron star (BNS) merger GW170817~\cite{gw170817}.
The inspiral part of the BNS signal in the Livingston detector coincided with a non-Gaussian noise transient which was caused by saturation in a digital-to-analog converter~\cite{gw170817}.
The initial LIGO-Virgo skymap for GW170817 became available only $\sim 4.5$ hours after the first LIGO-Virgo notice~\citep{gw170817_gcn_1, gw170817_gcn_2}.
While the amount of time required to remove a glitch has been reduced since the observation of GW170817, there was no tool in O3 that was able to remove a glitch in low latency, \textit{i.e.}\ within $\mathcal{O}$~(seconds-minutes).
Because glitches can affect parameter estimation of a GW event~\citep{glitch_pe_1, glitch_pe_2, glitch_pe_3, gw170817_bw}, including the sky localisation, such delay could negatively impact the efforts of low latency electromagnetic (EM) follow-up.
It is therefore crucial to understand what effect (if any) glitches have on GW skymaps, and how the localisation accuracy with different field-of-view (FOV) telescopes is affected.

In this paper we focus on estimating how non-Gaussian noise transients impact the sky localisation of a GW event in low-latency.
For the first time ever we asses the importance of relative positioning of a glitch with respect to a GW signal in time dimension, a task that became computationally possible only recently due to efficient GW search algorithms~\citep{pycbc_live, pycbc_live_o3}.
Section \S\ref{methods} describes the procedure we followed to mimic a real LIGO-Virgo search when a GW signal is in the vicinity of a glitch.
The section also discusses in more detail our glitch and GW signal samples.
In \S\ref{results} we present results of the study: how much sky localisation of specific GW events is affected by blip, thunderstorm and fast scattering glitches.
Section \S\ref{discussion} summarizes and discusses results, while conclusions are given in \S\ref{conclusions}.

\section{Methods}
\label{methods}

The main aim of the paper is to find how sky localisation of a GW signal is affected by non-Gaussian data in low latency.
We do this by mimicking a LIGO-Virgo low latency search in a scenario when a GW signal coincides with a noise transient.

To start with, we find a time where data in the Livingston interferometer contains a glitch whilst the other two detectors (Hanford and Virgo) have no data quality issues in a 1-minute window around the Livingston glitch time
(we explain in \S\ref{sub:injections} why we chose glitches in the Livingston detector).
Then we produce injections around this non-Gaussianity: a variety of GW signals are injected in three-(two-)detector data at a particular time relative to the glitch.
We repeat the same procedure at different times relative to the glitch in order to see how GW signal positioning relative to the glitch affects the recovered skymap.
We use the same strategy as used in O3 LIGO-Virgo low latency searches to detect a GW signal and provide a skymap for the event.
Finally, we assess what impact a coincident glitch with a GW signal could have on the EM follow-up campaign.
The exact procedure is outlined in the following subsections.

It is also worth nothing that this study and its parameter selection is mostly constrained by computational limits.
While extending the parameter space would have been possible (\textit{e.g}.~testing weaker GW signals), any additional results would give relatively small amount of new information comparing with a significant increase in the overall computational cost.

\subsection{Network selection}
\label{sub:network_selection}

We consider a network of three ground-based GW interferometers of LIGO Livingston, LIGO Hanford and Virgo.
We focus our study to three detectors for a number of reasons.
Firstly, we know that having a three-detector network greatly improves sky localisation accuracy~\cite{ifo_localization}, thus the effect of glitches should be easier to differentiate from random fluctuations caused by Gaussian-only noise.
In addition, the joint LIGO-Virgo three detector duty cycle reached $51\%$ in O3b~\cite{o3b_cat}, suggesting that for the majority of events we could expect data from three detectors.
However for completeness we also performed injections in two-detector LIGO network.
Finally, we omit KAGRA from our study because its sensitivity for the fourth observing run (O4) is planned to be at least an order of magnitude smaller than of Virgo or LIGO detectors~\cite{kagra_o4}.
Given these reasons, we believe that our network selection represents the GW network that would be available for O4. 

\subsection{Glitch sample}
\label{sub:glitches}

\begin{figure}
     \centering
     \begin{subfigure}[b]{0.4\paperwidth}
         \centering
         \includegraphics[width=0.4\paperwidth]{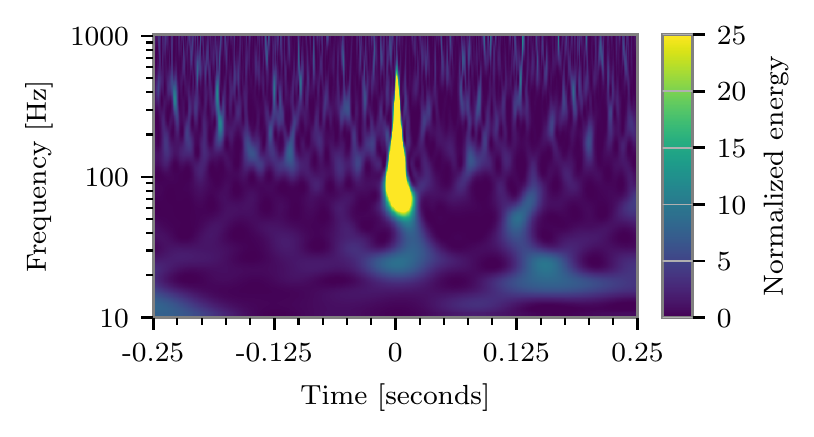}
         \caption{Blip gitch.}\label{fig:specgram_blip}
     \end{subfigure}
     \begin{subfigure}[b]{0.4\paperwidth}
         \centering
         \includegraphics[width=0.4\paperwidth]{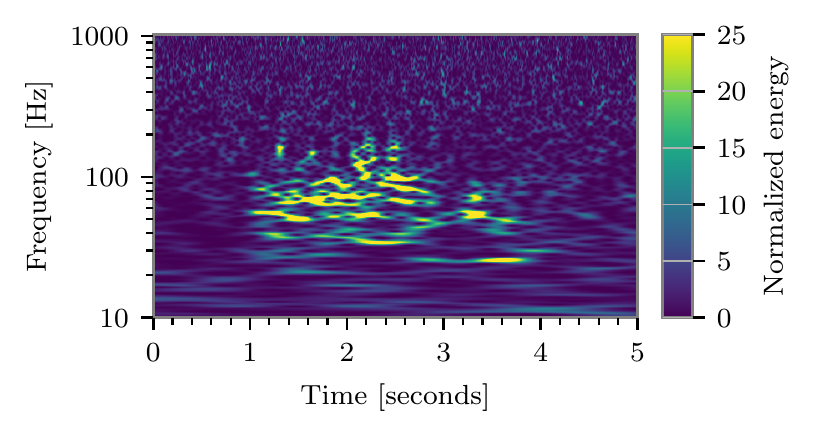}
         \caption{Thunderstorm glitch.}\label{fig:specgram_ts}
     \end{subfigure}
     \begin{subfigure}[b]{0.4\paperwidth}
         \includegraphics[width=0.4\paperwidth]{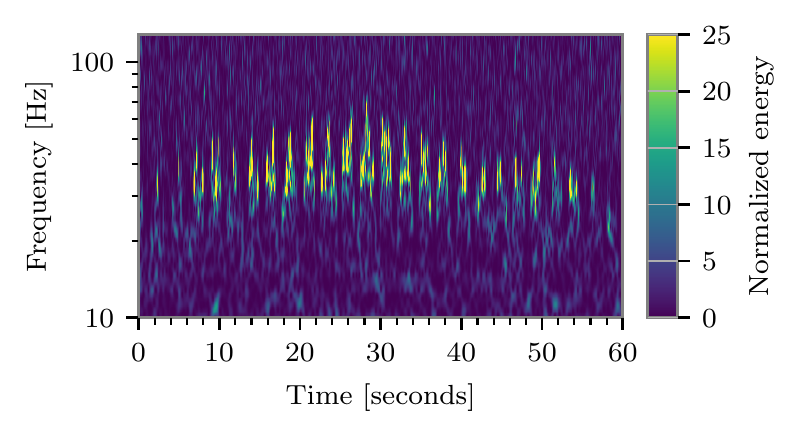}
         \caption{Fast scattering glitches.}\label{fig:specgram_sc}
     \end{subfigure}
        \caption{Time-frequency representations of three classes of glitches considered in our study.
            Blip glitches (a) are usually sub-second in duration and have a wide frequency bandwidth, $\mathcal{O}(100)$\,Hz.
            Thunderstorm glitches (b) are multiple-seconds ($3\text{-}10$\,s) in length and have a frequency of less than $200$\,Hz.
            Fast scattering glitches (c) are often found in groups and can last up to minutes with a typical frequency of $20\text{-}70$\,Hz.
            Note the different time and frequency axis values for each figure.
        }\label{fig:specgrams_glitches}
\end{figure}

In general there are many different glitches (and glitch types) in gravitational-wave interferometer data.
For example Gravity Spy, a glitch classification tool, has categorized approximately $3 \times 10^5$ glitches in 23 classes from O3 LIGO data alone~\cite{o3a_scattering}; it used machine learning and citizen science to classify glitches based on their time-frequency evolution~\cite{gspy}.
In this paper we chose to concentrate on three classes of glitches that have been observed in LIGO-Virgo data: blips, thunderstorms and fast scattering. 
Time-frequency representations of these glitches are shown in Figure~\ref{fig:specgrams_glitches}.

Blip glitches (Fig.~\ref{fig:specgram_blip}) are sub-second in duration and have a wide frequency bandwidth, $\mathcal{O}(100)$\,Hz~\cite{blips}.
It is currently unknown what causes most blip glitches in GW detectors, however in the second LIGO-Virgo observing run it was found that a subset of blip glitches were caused by computer timing errors~\cite{blips}. 
We chose blip glitches for our study because they affect high mass ($>100M_\odot)$ compact binary coalescence searches.
Blips are short, just like high mass binary black hole (BBH) events, hence they can be mistaken for a real GW event~\citep{blips, blips_imbh}.

Thunderstorm glitches (Fig.~\ref{fig:specgram_ts}), as the name suggests, are caused by thunderstorms which couple to the detector via acoustic noise~\cite{ts}.
They are usually $3\text{-}10$\,s in duration and have a frequency of less than $200$\,Hz.
While thunderstorm glitches are not as common as blip glitches, we chose them because they are within the most sensitive frequency range of GW interferometers ($60\text{-}200$\,Hz range).
Furthermore, thunderstorm glitches are in the middle of our time-frequency space between blips and scattering glitches (discussed later), \textit{i.e.}~the investigation of thunderstorm glitches should show how much localisation is affected by a wide frequency bandwidth glitch that lasts multiple seconds. 

Light scattering glitches are caused by stray light reflection in the beamtube, which can happen due to excessive ground motion~\cite{accadia_2010}.
For example, about $10\%$ of fast scattering glitches at Livingston during O3 were caused by trains passing by the GW interferometer~\cite{o3a_scattering}.
Light scattering glitches have characteristic arches that repeat over time which means that such noise transient can last up to minutes with a typical frequency range of $20\text{-}60$\,Hz (Fig.~\ref{fig:specgram_sc}).
Scattering glitches are sorted into two major categories, fast and slow~\cite{o3a_scattering}, depending on how often these arches repeat over time.
During O3, almost half of all glitches with signal-to-noise (SNR) ratio above 10 at the Livingston and Hanford interferometers were caused by light scattering ($44\%$ and $45\%$, respectively)~\cite{o3b_cat}. 
While slow scattering had been somewhat mitigated in O3b by reaction-chain tracking~\cite{o3b_cat}, fast scattering was still an issue.
In fact, it was the most frequent transient noise at Livingston during O3b~\cite{o3b_cat}.
Due to these reasons we selected fast scattering glitches (henceforth written as scattering glitches) for our study.

By considering these three classes of glitches we effectively cover the entire range of time-frequency noise that impacts transient GW searches the most.
We also expect these types of glitches to be some of the most prominent noise transients in O4.
There is nothing intrinsic about these glitches, \textit{i.e.}\ our conclusions for blip glitches should be applicable to other short duration glitches such as tomtes~\cite{o3a_scattering}, while results for thunderstorm and scattering glitches could be applied to any medium duration mid-frequency and long duration low-frequency glitches, respectively.

By using Gravity Spy~\cite{gspy}, we found 10 examples of each glitch type occuring at the Livingston detector in O3 data.
We assesed visually that data from Hanford and Virgo detectors did not have any noise noise artifacts for $\pm60$\,s around the Livingston glitch time $t_0$, while the data from Livingston detector was Gaussian-only at least $60$\,s before the Livingston glitch time $t_0$.

\subsection{GW injections}
\label{sub:injections}

With our sample of 30 blip, thunderstorm and scattering glitches, we injected simulated GW signals around each glitch.
Specifically, we inject the GW signal at a particular time relative to the glitch, and then investigate different times relative to the glitch to see if our results change depending on where a glitch coincides with the GW signal, \textit{e.g.}\ a blip glitch which overlaps the inspiral part of a BBH could have a different impact than a glitch overlapping the merger part.

In this paper we focused on three types of GW signals: BBHs, neutron star-black holes (NSBHs) and BNSs.
Given that BBH signals have a wide range of possible masses, we used three different events to represent a BBH-type signal.

All of our GW signals are based on real GW detections: GW190521, GW150914, GW170608, GW190814 and GW170817~\citep{gw190521, gw150914, gw170608, gw190814, gw170817}.
We used the \texttt{IMRPhenomD} waveform model~\cite{phenom_d} to generate BBH signals, \texttt{IMRPhenomNSBH} model~\cite{phenom_nsbh} to generate a NSBH signal and \texttt{IMRPhenomD\_NRTidalv2} model~\cite{phenom_nrtidal} to generate a BNS signal.
For simplicity we assumed that all our GW signals have zero spin.

Different signal types have distinctive durations and merging frequencies as outlined in Table~\ref{tab:signal_properties}.
Given a starting GW signal frequency of $20$\,Hz, all our tested BBH signals are shorter than $8$\,s.
The highest BBH merger frequency is for a GW170608-like event with $f_{merger}=835$\,Hz. 
A GW190814-like signal (representing a NSBH) is approximately $11$\,s duration with $f_{merger} = 458$\,Hz.
A GW170817-like signal (representing a BNS) is the longest signal in our set and has the highest merger frequency from all of our tested signals, $128$\,s and $4650$\,Hz, respectively.

\begin{table}
    \renewcommand{\arraystretch}{1.3}
    \begin{tabular}{p{1.5cm}p{2cm}p{1.5cm}p{1.2cm}p{1.0cm}} 
        \toprule
        Signal type & Event-like & Masses ($\text{M}_\odot$) & Length (s) & $f_{merger}$ (Hz) \\
        \midrule
        BBH & GW190521 & 85, 65 & 0.2 & 105 \\
        BBH & GW150914 & 36, 30 & 0.8 & 239 \\
        BBH & GW170608 & 11, 7.6 & 7.7 & 835 \\
        NSBH & GW190814 & 23, 2.6 & 11.4 & 531 \\
        BNS & GW170817 & 1.7, 1.7 & 128.0 & 1618 \\
        \bottomrule
    \end{tabular}
    \caption{Parameters of GW signals used in our study.
    We used three BBH events (GW190521-like, GW150914-like and GW170608-like), an NSBH event GW190814-like and a BNS event GW170817-like.
    The duration of GW signals is estimated assuming starting frequency of $20$\,Hz.
    Both parameters, duration in time and frequency, are important when considering the effect of glitches on GW signal localisation.
    }
    \label{tab:signal_properties}
\end{table}

GW signals are injected at a sky location to which Livingston detector has maximum sensitivity at the time of the glitch.
This is done because Livingston was the most sensitive detector in O3~\citep{o3a_cat, o3b_cat} and is projected to be the most sensitive detector in O4~\cite{kagra_o4}.
Furthermore, by fixing the sky location we can interpret our results more easily, as any changes in the calculated skymap will be due to the detector noise and not the location of the source.

If we were to instead choose a glitch in Hanford or Virgo rather than Livinsgton, the effect the glitch would have on the sky localisation bias would be smaller than in our tested scenario.
This is because we inject GW signals optimally for the Livingston detector (\textit{i.e.}~the highest SNR signal is in Livingston), thus there is more potential to bias results when a glitch overlaps a stronger GW signal than a weaker GW signal.

There is also another scenario to consider: what if a GW signal is not injected optimally at Livingston? 
In the majority of such cases Hanford and Virgo relative significance would become more important than in our tested scenario, making any effect of a Livingston glitch on the GW signal smaller than in our reported results.
This might not be true only if GW signals are injected in a such sky location where either Hanford or Virgo has no sensitivity.
This particular scenario needs additional studies.

Injected signals have SNR of $\{20, 30\}$ at the Livingston detector, where the SNR is defined as
\begin{equation}
    \label{eq:snr}
    \text{SNR} = \frac{\left|\left|\ \braket{s|h}\ \right|\right|}{\sqrt{\braket{h|h}}},
\end{equation}
where $s$ is strain data and $h$ is the gravitational-wave template.
The inner product is given by
\begin{equation}
    \label{eq:inner_product}
    \braket{a|b} = 4 \int_0^\infty \frac{\tilde{a}(f)\tilde{b}^*(f)}{S_n(f)}df,
\end{equation}
with $S_n(f)$ denotes the power spectral density (PSD).
In this analysis we used a PSD that was estimated at the glitch time for each detector individually.

The Livingston, Hanford and Virgo interferometers have different sensitivities.
During O3b the median BNS inspiral range for Livingston, Hanford and Virgo detectors was $133$\,Mpc, $115$\,Mpc and $51$\,Mpc, respectively~\cite{o3b_cat}.
This, together with the fact that each GW interferometer has a different sensitivity to a particular location of the sky, means that a GW signal injected optimally at the Livingston interferometer with $\text{SNR}_{\text{L1}}=\{20, 30\}$ corresponds to an injection at Hanford with $\text{SNR}_{\text{H1}} \approx \{16, 25\}$ and an injection at Virgo with $\text{SNR}_{\text{V1}} \approx\{8, 12\}$.
Combined total network SNR of such an injection is $\text{SNR}_{\text{net}} \approx \{27, 41\}$. 

For our two-detector injections we chose to use only $\text{SNR}_{L1}=20$.
Without the Virgo inteferometer the total network SNR is about 26.
Comparing with the three-detector $\text{SNR}_{L1}=20$ injections, the two-detector network $\text{SNR}_{Net}$ drops only by $4\%$, yet as we will see in~\S\ref{results} this has a significant impact.

Total network SNR and SNR of injections at the Hanford and Virgo detectors are given approximately because each detector's sensitivity varies throughout the observing run; an injection of $\textit{e.g.}$\, $\text{SNR}_{L1}=20$ corresponds to a range of SNRs at the Hanford interferometer with an approximate SNR of $16$.

We chose to make injections at $\text{SNR}_{L1}=\{20,30\}$ because in three-detector network $\text{SNR}_{L1}=20$ injections represent a weak three-detector detection (\textit{i.e.}\ $\text{SNR}_{V1}\approx8$ which is considered just above the detection threshold), while $\text{SNR}_{L1}=30$ represent a strong three-detector detection (\textit{i.e.}\ $\text{SNR}_{V1}\approx12$).

\subsection{Detection procedure}
\label{sub:detection_procedure}

Once GW signals are injected in LIGO-Virgo data containing a noise transient at Livingston, we follow the procedure used by low latency LIGO-Virgo searches in O3~\cite{pycbc_live_o3}.
To detect GW signals in low latency we employ \texttt{PyCBC Live}~\citep{pycbc_live, pycbc_live_o3}, an analysis package used to detect compact binary coalescences in close to real time.
In order to simulate a real search as close as possible, we used the same search parameters and analysis software version as in the LIGO-Virgo O3 low latency search.
For each GW event \texttt{PyCBC Live} returns various parameters, \textit{e.g.}~chirp mass~\cite{chirp_mass} and the SNR time-series.
We give this information to \texttt{BAYESTAR}, a rapid Bayesian position reconstruction package~\cite{bayestar}, which then computes a skymap of a GW event that can be investigated.

\subsection{Skymap interpretation}
\label{sub:skymap_interpretation}
For each glitch we made about 40 injections per GW signal per SNR\@. 
This means that our initial analysis contained approximately $12\,000$ skymaps.
It is not straightforward to interpret and compare such a large amount of skymaps, thus a skymap comparison quantity is needed that would be: (a) quantitative; (b) easily interpretable; (c) useful for astronomers.
We found that a tiling strategy~\cite{tiling_1, tiling_2, tiling_3} filled these requirements the best. 

Tiling is a method to scan the gravitational-wave sky localisation with a purpose to maximize the possibility of finding the corresponding EM event.
It divides a skymap into a telescope's FOV tiles and ranks them by contained skymap probability, thus instructing the telescope the order at which a skymap should be scanned. 
For our study we chose to adopt a tiling algorithm that was used by the Gravitational-wave Optical Transient Observer (GOTO) during O3~\citep{goto, goto_o3}, \texttt{GOTO-tile}~\cite{tiling_1}.
To simplify the comparison between different glitches, we chose to ignore such effects as the Sun and the Moon positioning at the time of telescope pointing, or the potential difference in the time required to point different FOV telescopes.

In this paper we focus on reporting tiling results, \textit{i.e.}\ which ranked tile contains the true sky location, for relatively small telescopes with a $\text{FOV} = 1\, \text{deg}^2$, as well as relatively large telescopes with a $\text{FOV} = 20\, \text{deg}^2$.
We also present in the paper another skymap comparison quantity, \textit{contour level}, which is independent on the telescope's FOV\@. 
Contour level corresponds to the skymap probability contour that contains the true GW event location.
It is common for astronomers to use $50\%$ or $90\%$ probability skymaps, thus we consider these contour values as important thresholds when determining if a glitch impacts the EM follow-up efforts. 

We report additional results in the Appendix~\ref{appendix}, such as tiling results for telescopes with FOVs of $0.25\,\text{deg}^2$, $5\,\text{deg}^2$, $10\,\text{deg}^2$ and $40\,\text{deg}^2$, $50\%$ and $90\%$ credible areas, and localisation distance.
Localisation distance is simply the angular distance between the true sky location of a GW event and the maximum probability pixel in the corresponding skymap.

\section{Results}
\label{results}

We present our results in three subsections where we discuss what effect blip (\S\ref{sub:blips}), thunderstorm (\S\ref{sub:ts}) and scattering (\S\ref{sub:sc}) glitches have on the localisation of our tested GW signals.
The way in which we present results is different for blips than for thunderstorm and scattering glitches.
Comparing with our tested GW signals (Table~\ref{tab:signal_properties}), the duration of a blip glitch can be essentially neglected.
This allows us to present how blip glitches affect the tiling efficiency simply by showing plots of the tiling efficiency change with respect to the glitch central time, $t_0$ (\textit{e.g.}\,Fig.~\ref{fig:gw150914_snr20_tiles_inner}).
For thunderstorm and scattering glitches, \textit{i.e.}~glitches with non-negligible duration, we rather show how tiling efficiency changes if the signal is \textit{within} the glitch versus \textit{outside} of the glitch (\textit{e.g.}\,Table~\ref{tab:ts}).

There is also additional complication due to the fact that thunderstorm and scattering glitches are extended over time.
Blip glitches, since they are so short, do not change their morphology appreciably over time, thus allowing us to average results from all 10 blip glitch runs.
This cannot be done for thunderstorm and scattering glitches, since they have extended duration, which means that their morphology can and does indeed change over time.
We found that results from 10 thunderstorm/scattering glitches vary greatly due to this, therefore we decided not to average results for these glitches and perform single-glitch representative runs instead.
Consequentially, our thunderstorm (\S\ref{sub:ts}) and scattering (\S\ref{sub:sc}) section discusses results from one respective glitch, while for blips (\S\ref{sub:blips}) we report averaged results from 10 blip glitch runs with $20\%$ trimming.
We trimmed ($\textit{i.e.}$\,removed) outliers that correspond to $20\%$ of all results because even random fluctuations in Gaussian-only noise can sometimes provide extreme values. 

Unless stated otherwise, `small' FOV refers to $1\, \text{deg}^2$ and `large' FOV refers to $20\, \text{deg}^2$.
For brevity, injections at $\text{SNR}_{Net}= 26$ refer to injections made at two-detector network of Livingston and Hanford with $\text{SNR}_{L1}=20$ and $\text{SNR}_{H1} \approx 16$.
Injections at $\text{SNR}_{Net}= \{27, 41\}$ refer to injections made with all three detectors with $\text{SNR}_{L1}=\{20, 30\}$, $\text{SNR}_{H1}\approx \{16, 25\}$ and $\text{SNR}_{V1}\approx\{8, 12\}$.

\subsection{Blips}
\label{sub:blips}

The rate of blip glitches at the Livingston and Hanford detectors during O3 was about 4 and 2 glitches per hour, respectively~\cite{blip_rate}.
Because blip glitches are about $\mathcal{O}(10)$\,ms in duration~\cite{blips}, only in rare cases would we expect a blip glitch to overlap a GW signal.

\subsubsection{BBH}
\label{ssub:blips_bbh}
\begin{figure}[htp]
    \centering
    \includegraphics[width=\columnwidth, trim={0 0 0 0cm},clip]{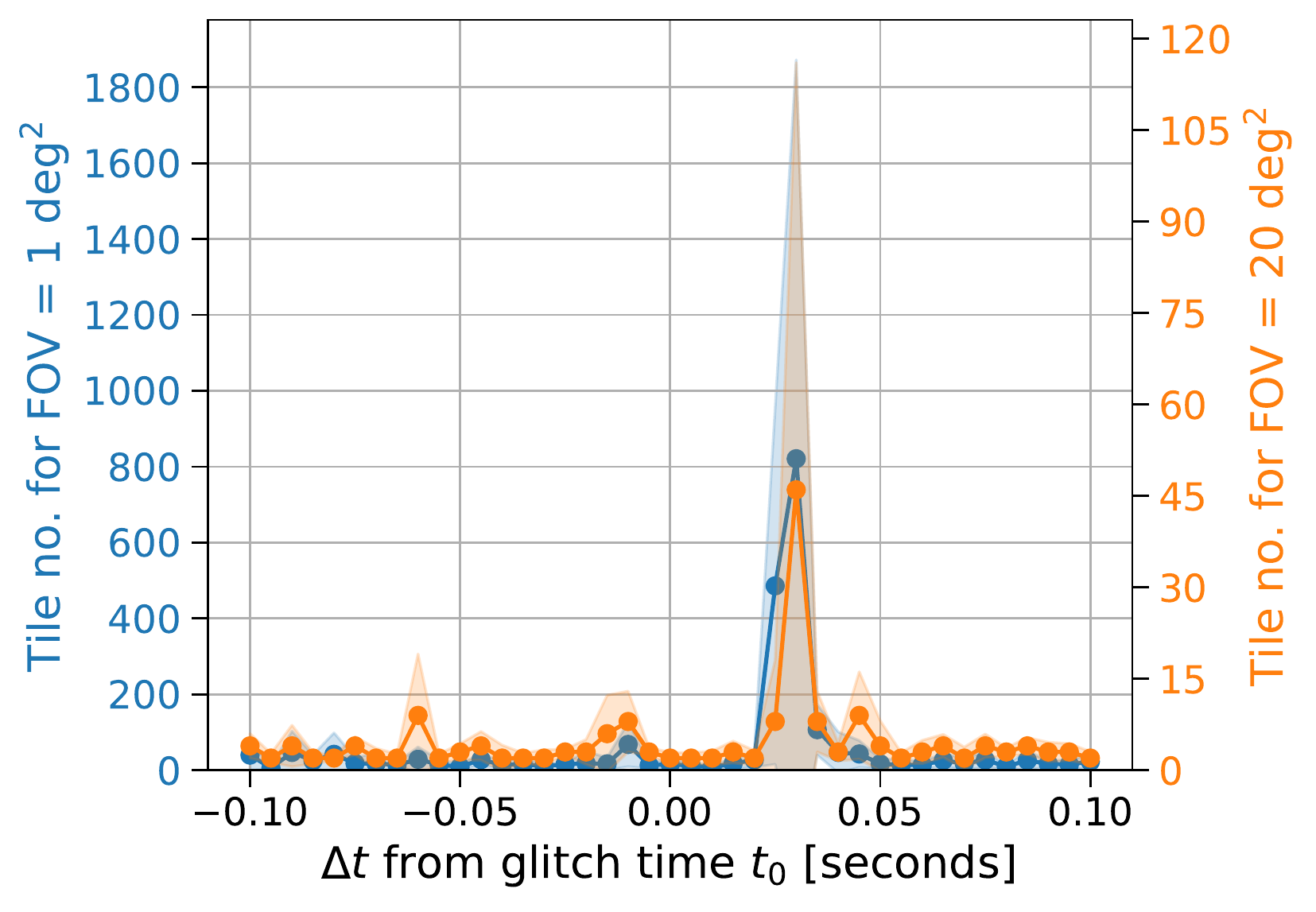}
    \caption{Average number of tiles required to observe the true sky location of a GW150914-like $\text{SNR}_{Net}=27$ event relative to the blip glitch central time $t_0$. 
    For both FOV telescopes, small ($1\,\text{deg}^2$, orange) and large ($20\,\text{deg}^2$, blue), tiling deficiency at $t_0+30$\,ms is observed.
    Shaded bands represent $1\sigma$ deviation.
    Note that these are averaged results from $10$ blip glitches with $20\%$ trimming.
    }
    \label{fig:gw150914_snr20_tiles_inner}
\end{figure}
\begin{figure}[htp]
    \centering
    \includegraphics[width=\columnwidth, trim={0 0 0cm 0cm},clip]{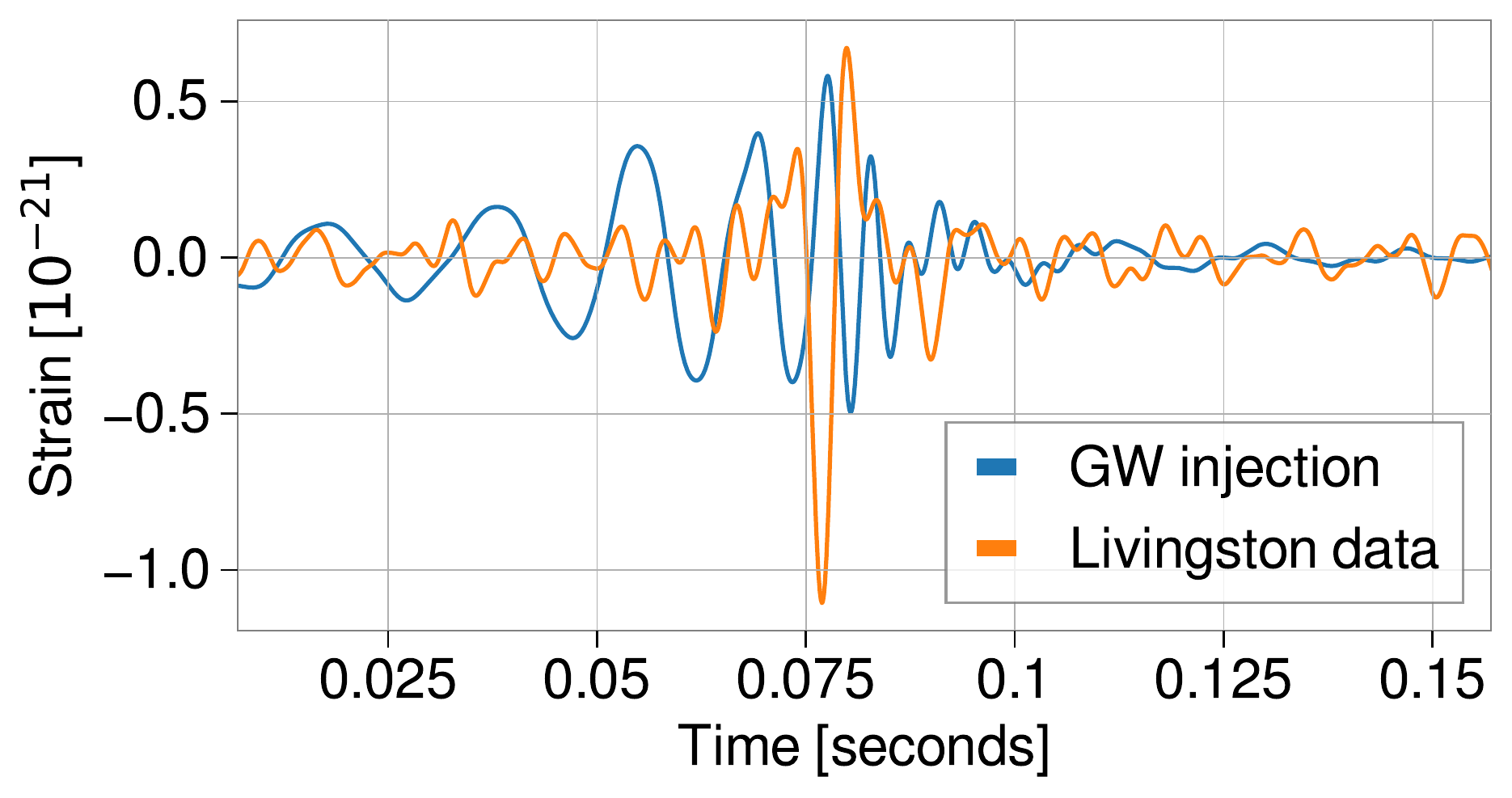}
    \caption{Time series data of a GW150914-like injected signal (blue) in Livingston data (orange) containing a blip glitch.
    The GW signal is injected at $t_0+30$\,ms relative to the blip glitch central time $t_0$.
    The glitch is aligned with the peak GW signal in a such way that they have opposite phases.
    }\label{fig:tseries_phase_orig}
\end{figure}
\begin{figure}
     \centering
     \begin{subfigure}[b]{0.4\paperwidth}
        \centering
        \includegraphics[width=\columnwidth, trim={0 5.5cm 0 5.5cm},clip]{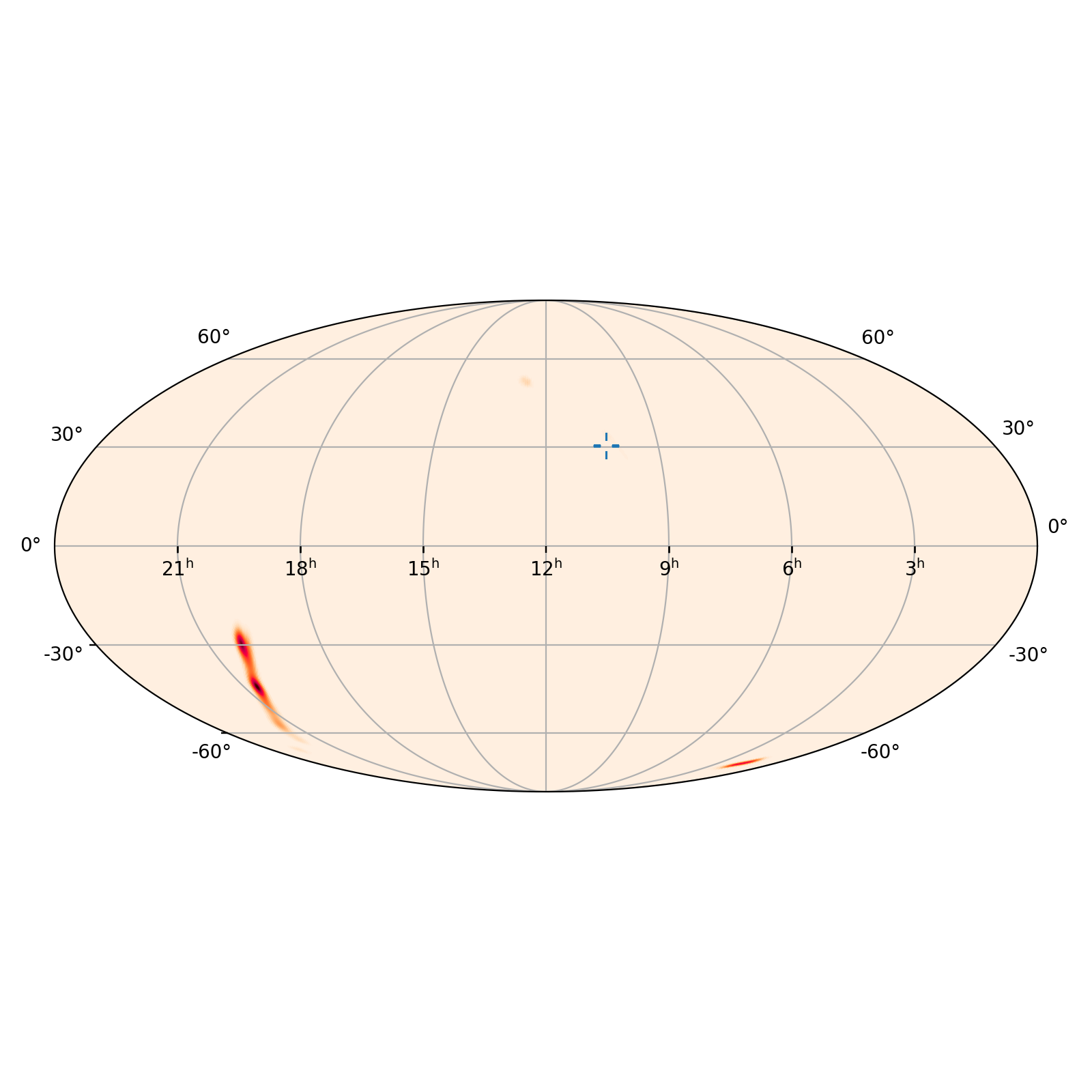}
        \caption{Sky localisation of a GW150914-like event injected at $t_0+30$\,ms relative to the blip glitch central time $t_0$. The $90\%$ credible area is $137\,\text{deg}^2$.}
        \label{fig:skymap_orig}
     \end{subfigure}
     \begin{subfigure}[b]{0.4\paperwidth}
        \centering
        \includegraphics[width=\columnwidth, trim={0 5.5cm 0 5.0cm},clip]{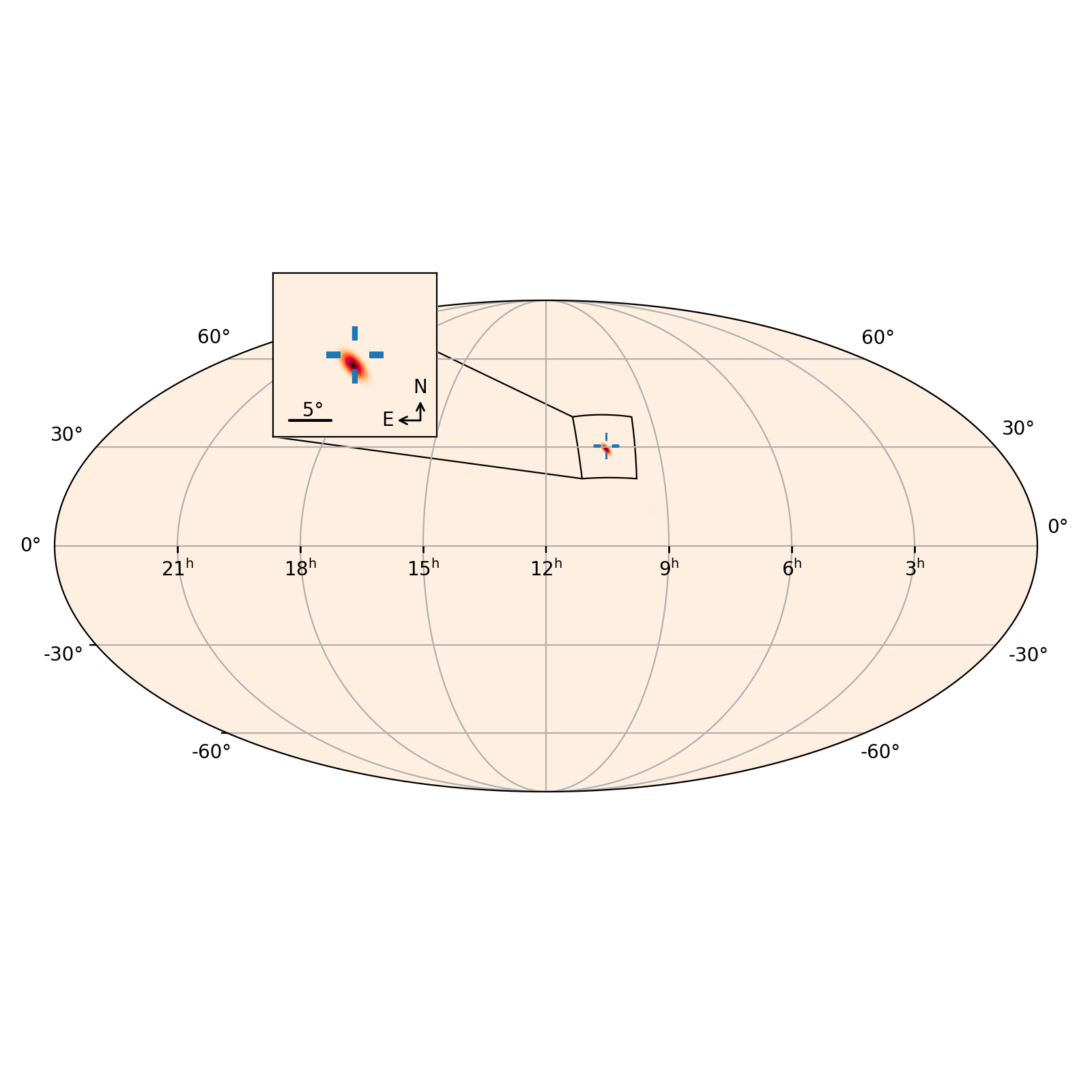}
         \caption{Identical skymap to Fig.~\ref{fig:skymap_orig} except that the injected GW signal was phase-shifted by $\pi/2$. The $90\%$ credible area is $8\,\text{deg}^2$.}
        \label{fig:skymap_inv}
     \end{subfigure}
     \caption{Sky localisation of a GW150914-like event injected at $t_0+30$\,ms relative to the blip glitch central time $t_0$.
     The skymap of the time series data from Fig.~\ref{fig:tseries_phase_orig} (top) shows that the true sky location of the event (indicated by blue pointer) does not coincide with the estimated sky location.
     This happens because the glitch overlaps and cancels part of the GW signal.
     If the same GW signal is phase-shifted by $\pi/2$, the true event sky location agrees well with the estimated sky location (bottom).
     }
     \label{fig:gw150914_skymaps}
\end{figure}

Given the duration and merging frequencies of our tested BBH signals (Table~\ref{tab:signal_properties}), we decided to use a conservative window of $\pm0.5$\,s around the central glitch time $t_0$ to inject GW signals.
Signals were injected with $50$\,ms spacing, however closer to the glitch $t_0$ ($\pm0.1$\,s around the glitch $t_0$) we used a finer spacing of $5$\,ms.
We found that the tiling and contour level results do not vary noticeably for any of the BBH signals at any times with one particular exception.
All of our tested BBH signals have worse tiling results (\textit{i.e.}~a tiling deficiency) around $30$\,ms after the glitch time $t_{0}$.
As an example, see Figure~\ref{fig:gw150914_snr20_tiles_inner} for a GW150914-like signal with $\text{SNR}_{Net}=27$.

Further investigation revealed that such a tiling bias occurs only if the blip glitch destructively interferes with the GW waveform close to the merger time (Fig.~\ref{fig:tseries_phase_orig}).
This is possible only if a blip glitch has the opposite phase to the GW waveform at a particular time relative to the glitch (in this case it is $t_{0}+30$\,ms).

Considering all blip glitch runs, we find that the number of tiles required to search over to find the true sky location of an event can change significantly.
In the worst-case scenario the number of tiles that need to be searched over increases by a factor of over $500$\,($180$) for a FOV=$1\,\text{deg}^2$ ($20\,\text{deg}^2$) telescope for an event like GW190521.
For a GW150914-like event it increases by a factor of over $850$\,($400$) for a FOV=$1\,\text{deg}^2$ ($20\,\text{deg}^2$).
For a GW170608-like event it increases by a factor of $368$\,($90$) for a FOV=$1\,\text{deg}^2$ ($20\,\text{deg}^2$).

In some cases the $90\%$ credible area can also change drastically.
Figure~\ref{fig:gw150914_skymaps} shows skymaps of a GW150914-like event at $t_0+30$\,ms relative to the glitch time.
Figure~\ref{fig:skymap_orig} presents the corresponding skymap of time-series from Fig.~\ref{fig:tseries_phase_orig} with the $90\%$ credible area of $137\,\text{deg}^2$.
Figure~\ref{fig:skymap_inv} skymap shows the localisation of the same time-series as in Fig.~\ref{fig:tseries_phase_orig}, but with the shifted GW injection phase by $\pi/2$.
In this case the $90\%$ credible area is reduced to $8\,\text{deg}^2$ and overlaps with the true sky location.

\subsubsection{NSBH}
\label{ssub:blips_nsbh}

For a GW190814-like signal we performed injections in $[-1,+9]$\,s window around the blip glitch time $t_0$ with injection spacing of $300$\,ms.
We extended the injection window from $[-0.5, +0.5]$\,s because a GW190814-like signal is longer than any of our tested BBH signals.
Closer to the glitch, \textit{i.e.}~$[-0.025, +0.1]$\,s around the glitch $t_0$, a finer spacing of $5$\,ms was used.

\begin{figure}
     \centering
        \includegraphics[width=\columnwidth]{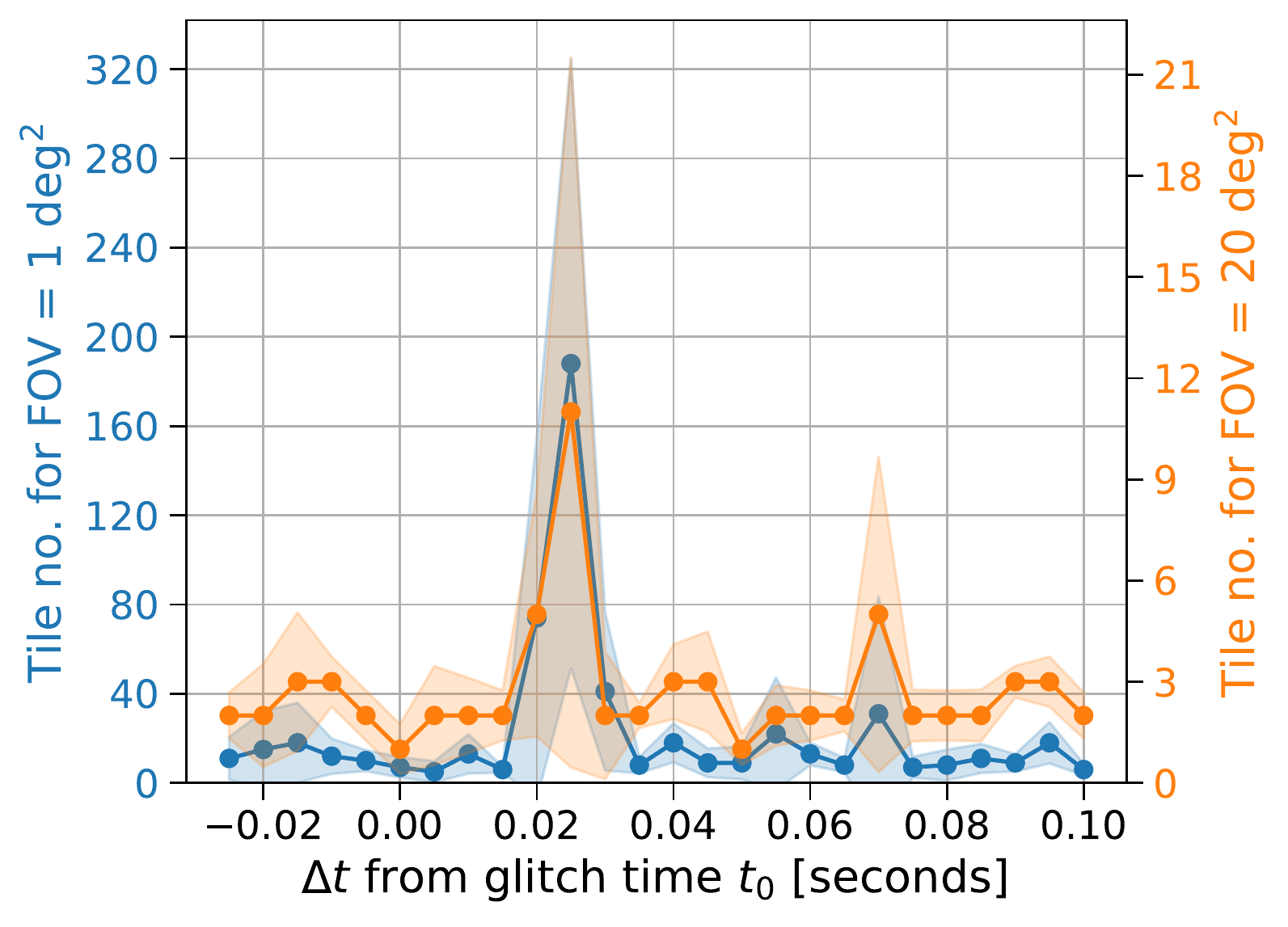}
     \caption{Average number of tiles required to observe the true sky location of a GW190814-like event for $\text{SNR}_{Net}=27$.
         For both FOV telescopes, small ($1\,\text{deg}^2$, orange) and large ($20\,\text{deg}^2$, blue), the tiling deficiency at $t_0+25$\,ms is observed.
         Shaded bands represent $1\sigma$ deviation.
         Note that these are averaged results from $10$ blip glitches with $20\%$ trimming.
     }
     \label{fig:gw190814_tiles_inner}
\end{figure}

Similarly to the BBH-type signals, a GW190814-like event has a significant change in tiling results only at one place: about $25$\,ms after the glitch $t_0$ (Figure~\ref{fig:gw190814_tiles_inner}).

In the worst-case scenario, $106$ ($17$) times more tiles are required to find the true sky location of an event for FOV=$1\,\text{deg}^2$ ($20\,\text{deg}^2$) telescopes.
We did not find any noticeable difference in the $90\%$ credible area for GW190814-like events.

\subsubsection{BNS}
\label{ssub:blips_bns}
For a GW170817-like signal we used identical injection window parameters as in the NSBH case: injection window of $[-1, +9]$\,s around the glitch time $t_0$ with injections at $300$\,ms intervals, and the inner window of $[-0.025, +0.1]$\,s around the glitch $t_0$ with injections at $5$\,ms intervals.

\begin{figure}[htp]
    \centering
    \includegraphics[width=\columnwidth]{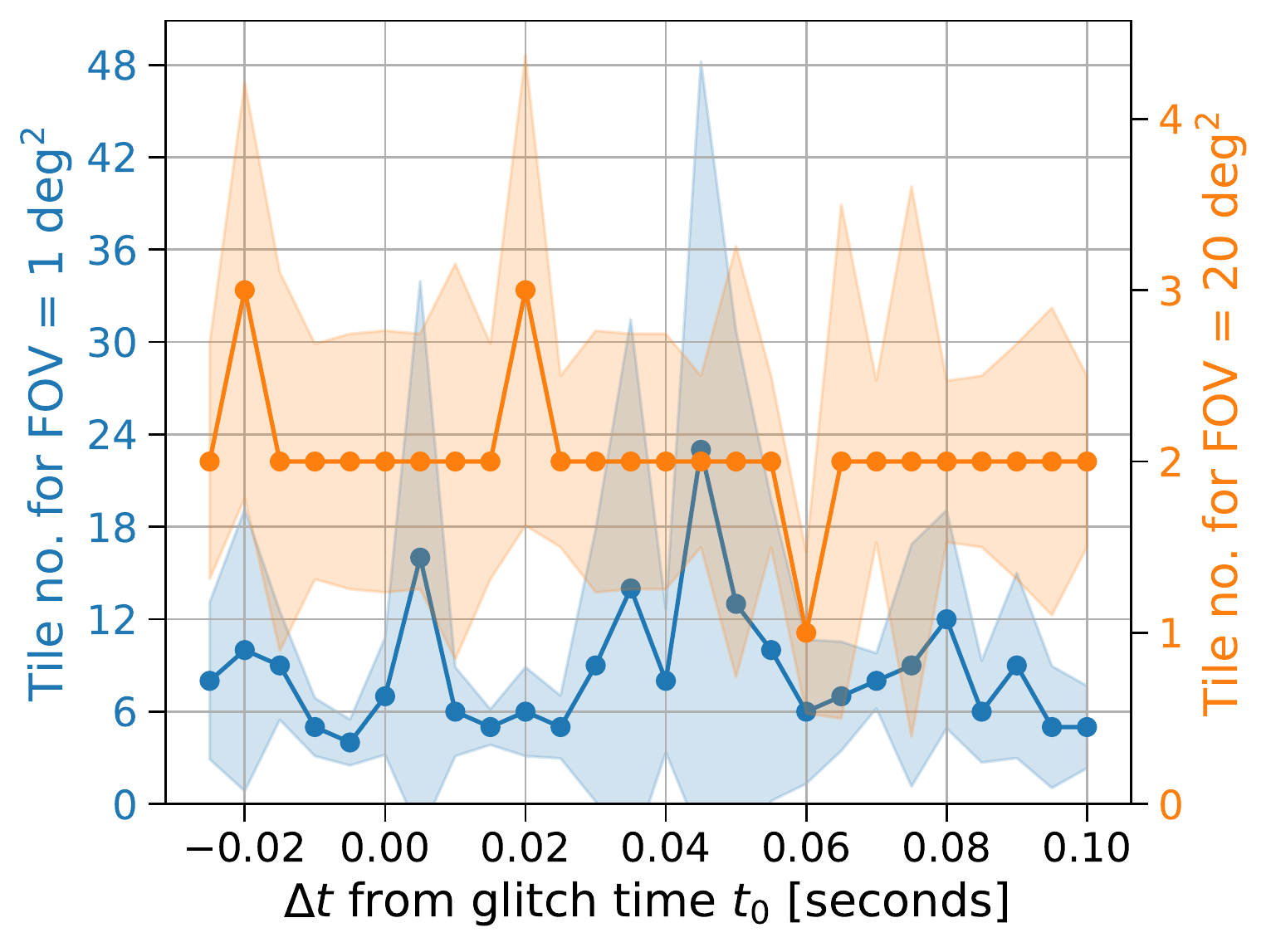}
    \caption{Average number of tiles required to observe the true sky location of a GW170817-like $\text{SNR}_{Net}=27$ event relative to the blip glitch central time $t_0$. 
    There is no noticeable change in the tiling efficiency for both FOV telescopes, small ($1\,\text{deg}^2$, orange) and large ($20\,\text{deg}^2$, blue).
    Shaded bands represent $1\sigma$ deviation.
    Note that these are averaged results from $10$ blip glitches with $20\%$ trimming.
    }
    \label{fig:gw170817_snr20_tiles_inner}
\end{figure}

Our results indicate that a BNS tiling efficiency for small and large FOV telescopes is not affected by blip glitches (Fig.~\ref{fig:gw170817_snr20_tiles_inner}).
We argue that this happens because a BNS signal at the merger part has a much higher frequency and smaller amplitude than a NSBH or BBH signal (Table~\ref{tab:signal_properties}), thus destructive interference with a blip glitch like in Fig.~\ref{fig:tseries_phase_orig} is unlikely to occur.

\subsection{Thunderstorm glitches}
\label{sub:ts}

Typical thunderstorm glitches are about $3\text{-}10$\,s duration and have a frequency of less than $200$\,Hz.
Given these parameters, we expect that BBH, NSBH and BNS signals should overlap with such glitches in frequency and thus could be affected.

We report contour and tiling results for GW signals in Table~\ref{tab:ts}.
Additional results for thunderstorm glitches can be found in the Appendix (Table~\ref{tab:ts_otherFOVs}).
The `Pre-glitch' column in Table~\ref{tab:ts} and Table~\ref{tab:ts_otherFOVs} refers to results from injections before the glitch start time $t_0$, while the `Glitch' column refers to results from injections after the glitch $t_0$. 

\begin{table*}
    \renewcommand{\arraystretch}{1.3}
    \begin{tabular}{p{2.3cm}p{3.3cm}p{1.6cm}p{1.6cm}p{0.3cm}p{1.6cm}p{1.6cm}p{0.3cm}p{1.6cm}p{1.6cm}} 
        \toprule
          & & \multicolumn{2}{c@{}}{$\text{SNR}_{Net}=26$ (2 detectors)} & & \multicolumn{2}{c@{}}{$\text{SNR}_{Net}=27$} & & \multicolumn{2}{c@{}}{$\text{SNR}_{Net}=41$} \\
        \cmidrule(l){3-4} \cmidrule(l){6-7} \cmidrule(l){9-10} 
          & & Pre-glitch & Glitch & & Pre-glitch & Glitch & & Pre-glitch & Glitch \\ 
        \midrule
        GW190521-like & Contour level & 0.16 $\pm$ 0.12 & 0.26 $\pm$ 0.21 && 0.34 $\pm$ 0.14 & 0.43 $\pm$ 0.16  & & 0.34 $\pm$ 0.15  &  0.46 $\pm$ 0.17 \\
                          & Tile no.~(FOV = $1\,\text{deg}^2$) & 60 $^{+61}_{-59}$ & 138 $^{+141}_{-137}$ && 10 $\pm$ 5 &  15 $\pm$ 10  &  & 4 $\pm$ 2  & 6 $\pm$ 3 \\
                          & Tile no.~(FOV = $20\,\text{deg}^2$)& 6 $\pm$ 4 & 11 $\pm$ 9 && 2 $\pm$ 1  & 2 $\pm$ 1 & & 1 $^{+1}_{-0}$  & 1 $^{+1}_{-0}$ \\
                          \midrule
        GW150914-like & Contour level & 0.14 $\pm$ 0.10 & 0.23 $\pm$ 0.18 && 0.35 $\pm$ 0.15  & 0.36 $\pm$ 0.15  & & 0.39 $\pm$ 0.15  &  0.40 $\pm$ 0.17 \\
                          & Tile no.~(FOV = $1\,\text{deg}^2$) & 38 $\pm$ 34 & 73 $\pm$ 69 && 13 $\pm$  9 & 13 $\pm$ 9 & & 4 $\pm$ 2  & 4 $\pm$ 2 \\
                          & Tile no.~(FOV = $20\,\text{deg}^2$)& 5 $\pm$ 4 & 9 $\pm$ 7 && 2 $\pm$ 1 & 2 $\pm$ 1 & & 1 $^{+1}_{-0}$  & 1 $^{+1}_{-0}$ \\
                          \midrule
        GW170608-like & Contour level & 0.25 $\pm$ 0.17 & 0.28 $\pm$ 0.17 && 0.42 $\pm$ 0.16  & 0.41 $\pm$ 0.17  & & 0.42 $\pm$ 0.15  & 0.40 $\pm$ 0.17 \\
                          & Tile no.~(FOV = $1\,\text{deg}^2$)  & 33 $\pm$ 28 & 44 $\pm$ 39 && 6 $\pm$ 3 & 6 $\pm$ 3 & & 2 $\pm$ 1  & 2 $\pm$ 1 \\
                          & Tile no.~(FOV = $20\,\text{deg}^2$) & 5 $\pm$ 4 & 7 $\pm$ 5 && 2 $\pm$ 1 & 2 $\pm$  1 & & 1 $\pm$ 0  & 1 $\pm$ 0 \\
                          \midrule
        GW190814-like & Contour level & 0.23 $^{+0.27}_{-0.23}$ & 0.51 $\pm$ 0.33 && 0.32 $\pm$ 0.26 & 0.49 $\pm$ 0.29 & & 0.35 $\pm$ 0.27  & 0.50 $\pm$ 0.28 \\
                          & Tile no.~(FOV = $1\,\text{deg}^2$)  & 75 $^{+156}_{-74}$ & 278 $^{+376}_{-277}$ && 22 $^{+69}_{-21}$ & 43 $^{+108}_{-42}$ & & 3 $^{+3}_{-2}$  & 6 $^{+7}_{-5}$ \\
                          & Tile no.~(FOV = $20\,\text{deg}^2$) & 13 $^{+13}_{-12}$ & 32 $^{+36}_{-31}$ && 4 $^{+6}_{-3}$ & 6 $^{+11}_{-5}$ & & 2 $\pm$ 1  & 2 $\pm$ 1 \\
                          \midrule
        GW170817-like & Contour level & 0.25 $^{+0.28}_{-0.25}$ & 0.26 $\pm$ 0.26 && 0.46 $\pm$ 0.30  & 0.41 $\pm$ 0.30  & & 0.41 $\pm$ 0.27  & 0.39 $\pm$ 0.26 \\
                          & Tile no.~(FOV = $1\,\text{deg}^2$)  & 49 $^{+84}_{-48}$ & 52 $^{+84}_{-51}$ && 28 $^{+69}_{-27}$ & 25 $^{+65}_{-24}$ &  & 3 $\pm$ 2  & 3 $^{+4}_{-2}$ \\
                          & Tile no.~(FOV = $20\,\text{deg}^2$) & 11 $\pm$ 9 & 13 $\pm$ 10 && 5 $^{+8}_{-4}$ & 5 $^{+8}_{-4}$ & & 2 $\pm$ 1  & 2 $^{+2}_{-1}$ \\
        \bottomrule
    \end{tabular}
    \caption{Tiling and contour level results for BBH, NSBH and BNS type signals, for the thunderstorm glitch injection study.
    $\text{SNR}_{Net}=26$ corresponds to results from injections made only in Livingson and Hanford, \textit{i.e.}\,no Virgo, while $\text{SNR}_{Net}=27$ and $\text{SNR}_{Net}=41$ refer to results with injections made in all three interferometers.
    Tiling results report the number of tile pointings that a telescope would need to search over until the true sky location of an event is observed.
    Contour level corresponds to the skymap probability contour that contains the true GW event location.
    The `Pre-glitch' column refers to injections that were made before the thunderstorm glitch start time $t_0$ (this represents an injection sample that is not affected by the glitch).
    The `Glitch' column refers to injections that were made after the thunderstorm glitch start time $t_0$ (this represents an injection sample that is affected by the glitch).
    The table reports averaged results from the `Pre-glitch' and `Glitch' samples with $1\sigma$ deviation.
    For BBH type signals these are averaged results from $10$ glitches with $20\%$ trimming, while NSBH and BNS type results are from a single representative glitch.
    This also explains why BBH results have generally smaller error bars than NSBH and BNS results.
    }
    \label{tab:ts}
\end{table*}

\subsubsection{BBH}
\label{ssub:ts_bbh}

While thunderstorm glitches can have different durations, we noted that many thunderstorm glitches look similar at the start of a glitch, \textit{i.e.}\ they have a distinctively loud wide-frequency excess power ($1\text{-}3$\,s in Fig.~\ref{fig:specgram_ts}).
This distinctive part of a thunderstorm glitch usually contains more power and has a wider frequency content than the rest of the glitch, thus we consider it as the worst part of the glitch.  
Consequently, we decided to inject BBH signals $\pm2.5$\,s around the thunderstorm glitch start time, $t_0$, with the injection spacing of $100$\,ms.

All our reported thunderstorm glitch results for BBHs are averaged from 10 glitch runs with $20\%$ trimming; this was done in order to reduce statistics skewing due to random fluctuations in Gaussian noise.

We found that GW190521-type signals are impacted the most from all BBH signals.
Results are significantly different between `Pre-glitch' and `Glitch' windows for two detector injections.
Tile number increases by about $100\%$ for both, small and large FOV telescopes.
In the presence of a thunderstorm glitch the tile \#138 instead of tile \#60 contains the true sky location for FOV = $1\,\text{deg}^2$ telescope.
For a telescope with a FOV = $20\,\text{deg}^2$ tile \#11 instead of tile \#6 contains the true sky location.
The contour level also increases from 0.16 to 0.26.

For three detector injections, the impact of a thunderstorm glitch is such that only the small FOV telescopes are affected.
For both SNRs, $\text{SNR}_{Net}=27$ and $\text{SNR}_{Net}=41$, the number of tiles that need to be observed with FOV=$1\,\text{deg}^2$ telescope increase by about $50\%$: 
tile \#15 instead of tile \#10 for $\text{SNR}_{Net}=27$ and tile \#6 instead of tile \#4 for $\text{SNR}_{Net}=41$.
The contour level increased by about $25\%$ but is still within the 50th percentile.

Lower mass black hole events, such as GW150914-like and GW170608-like, appear to be impacted by thunderstorm glitches only if there is no third detector, \textit{i.e.}\,in the $\text{SNR}_{Net}=26$ case.
Both, small and large FOV telescopes are affected, while the contour level somewhat changes but does not exceed $50\%$ (Table~\ref{tab:ts}).

\subsubsection{NSBH}
\label{ssub:ts_nsbh}

Similarly to the BBH injections, we injected a GW190814-like signal around 10 thunderstorm glitches.
Because a GW190814-like event is much longer than any of our tested BBH signals, we extended our injection window to $[t_0-1, t_0 + \text{glitch duration} + 9]$\,s.
However we quickly noticed that averaging results from 10 thunderstorm glitches becomes impractical because the morphology of thunderstorm glitches varies a lot.

Instead of averaging results from multiple glitches, we decided to perform a single representative run on one of the strongest thunderstorm glitches from our study.
For this glitch we made injections $\pm3$\,s around the glitch start time, $t_0$, with the injection spacing of $50$\,ms.

We notice significant changes in tiling results almost in all cases.
For small FOV telescopes at $\text{SNR}_{Net}=26$, there is almost fourfold increase in the number of tiles needed to be searched over (tile \#278 instead of tile \#75).
For three detector injections the difference becomes smaller, about twofold on average, for both, $\text{SNR}_{Net}=27$ and $\text{SNR}_{Net}=41$.
The localisation with large FOV telescopes is impacted with $\text{SNR}_{Net}=26$ and $\text{SNR}_{Net}=27$ configuration but at high SNR, \textit{i.e.}\,$\text{SNR}_{Net}=41$, there are no differences between `Pre-glitch' and `Glitch' injections (Table~\ref{tab:ts}).

The impact of thunderstorm glitches for NSBH sources is also observed in contour level.
For two and three detector injections the contour level increases and sometimes exceeds the 50th percentile.

\subsubsection{BNS}
\label{ssub:ts_bns}

We used an injection window of $[t_0-1, t_0 + \text{glitch duration} + 9]$\,s for a GW170817-like event.
As with the NSBH case (\S\ref{ssub:ts_nsbh}), we found that averaging results from 10 thunderstorm glitches does not make sense because of the variation in the morphology of thunderstorm glitches.
For that reason we performed injections on the same thunderstorm glitch as in the NSBH case.
Since a GW170817 signal is longer than GW190814, an appropriately larger injection window was chosen: $\pm8$\,s around the glitch the thunderstorm glitch start time, $t_0$, with the injection spacing of $100$\,ms.

Results for two detector ($\text{SNR}_{Net}=26$) and three detector ($\text{SNR}_{Net}=27$ and $\text{SNR}_{Net}=41$) injections indicate that there are no significant changes in contour level and tiling efficiency for both, small and large FOV telescopes (Table~\ref{tab:ts}).

\subsection{Scattering glitches}
\label{sub:sc}

Out of all three types of glitches we tested, scattering glitches have the most complicated morphology.
Scattering glitches (Fig.~\ref{fig:specgram_sc}) can last hundreds of seconds, often with varying frequency, which means that any generalisation of such glitches is non-trivial.

Scattering glitches are within the lower frequency range of a ground-based GW detector sensitivity ($<100$\,Hz), therefore we expect such glitches to overlap with all of our tested signal types, \textit{i.e.}\ BBHs, NSBHs and BNSs. 

We report the tiling and contour level results for GW signals in Table~\ref{tab:sc}.
Additional results for scattering glitches can be found in the Appendix (Table~\ref{tab:sc_otherFOVs}).
The `Pre-glitch' column in Table~\ref{tab:sc} and Table~\ref{tab:sc_otherFOVs} refers to results from injections some time before the scattering glitch start, while the `Glitch' column refers to results from injections during the glitch.

\begin{table*}
    \renewcommand{\arraystretch}{1.3}
    \begin{tabular}{p{2.3cm}p{3.3cm}p{1.6cm}p{1.6cm}p{0.3cm}p{1.6cm}p{1.6cm}p{0.3cm}p{1.6cm}p{1.6cm}} 
        \toprule
         & & \multicolumn{2}{c@{}}{$\text{SNR}_{Net}=26$ (2 detectors)} & &\multicolumn{2}{c@{}}{$\text{SNR}_{Net}=27$} & & \multicolumn{2}{c@{}}{$\text{SNR}_{Net}=41$} \\
        \cmidrule(l){3-4} \cmidrule(l){6-7} \cmidrule(l){9-10} 
          &  & Pre-glitch & Glitch & & Pre-glitch & Glitch & & Pre-glitch & Glitch \\ 
        \midrule
        GW190521-like & Contour level & 0.15 $\pm$ 0.10 & 0.16 $\pm$ 0.13  && 0.41 $\pm$ 0.17  & 0.39 $\pm$ 0.16  & & 0.42 $\pm$ 0.16 & 0.38 $\pm$ 0.14 \\
                          & Tile no.~(FOV = $1\,\text{deg}^2$)  & 58 $\pm$ 53 & 63 $\pm$ 61 && 14 $\pm$ 9 &  13 $\pm$ 8 & & 6 $\pm$ 3 & 5 $\pm$ 2 \\
                          & Tile no.~(FOV = $20\,\text{deg}^2$) & 8 $^{+8}_{-7}$ & 9 $\pm$ 8 && 2 $\pm$ 1 & 2 $\pm$ 1 & & 2 $\pm$ 1 & 1 $^{+1}_{-0}$ \\
        \midrule
        GW150914-like & Contour level & 0.18 $\pm$ 0.16 & 0.16 $\pm$ 0.14 && 0.43 $\pm$ 0.18  & 0.33 $\pm$ 0.15  & & 0.44 $\pm$ 0.18  & 0.36 $\pm$ 0.15 \\
                          & Tile no.~(FOV = $1\,\text{deg}^2$)  & 36 $^{+42}_{-35}$ & 36 $^{+39}_{-35}$ && 7 $\pm$ 6 & 5 $\pm$ 3  & & 3 $\pm$ 2 & 3 $\pm$ 1 \\
                          & Tile no.~(FOV = $20\,\text{deg}^2$) & 7 $\pm$ 6 & 8 $\pm$ 5 && 2 $\pm$ 1  & 2 $\pm$ 1  & &  1 $^{+1}_{-0}$ & 1 $\pm$ 0 \\
        \midrule
        GW170608-like & Contour level & 0.31 $\pm$ 0.20 & 0.26 $\pm$ 0.17 && 0.49 $\pm$ 0.19  & 0.39 $\pm$ 0.15  & & 0.52 $\pm$ 0.18 & 0.43 $\pm$ 0.15 \\
                          & Tile no.~(FOV = $1\,\text{deg}^2$)  & 53 $\pm$ 50 & 43 $\pm$ 40 && 8 $\pm$ 6 & 5 $\pm$ 3 & & 3 $\pm$ 2 & 2 $\pm$ 1  \\
                          & Tile no.~(FOV = $20\,\text{deg}^2$) & 9 $\pm$ 6 & 8 $\pm$ 5 && 2 $\pm$ 1 & 2 $\pm$ 1 & & 1 $^{+1}_{-0}$ & 1 $\pm$ 0 \\
        \midrule
        GW190814-like & Contour level & 0.17 $^{+0.20}_{-0.17}$ & 0.20 $^{+0.26}_{-0.20}$ && 0.35 $\pm$ 0.27  & 0.39 $\pm$ 0.24  & & 0.35 $\pm$ 0.24 & 0.41 $\pm$ 0.28 \\
                          & Tile no.~(FOV = $1\,\text{deg}^2$)  & 37 $^{+60}_{-36}$ & 95 $^{+240}_{-94}$ && 10 $^{+16}_{-9}$ & 17 $^{+33}_{-16}$ & & 3 $^{+3}_{-2}$ & 7 $^{+24}_{-6}$ \\
                          & Tile no.~(FOV = $20\,\text{deg}^2$) & 6 $\pm$ 5 & 13 $^{+19}_{-12}$ && 2 $\pm$ 1 & 3 $^{+3}_{-2}$ & & 1 $\pm$ 0 & 2 $^{+3}_{-1}$ \\
        \midrule
        GW170817-like & Contour level & 0.23 $^{+0.27}_{-0.23}$ & 0.23 $^{+0.26}_{-0.23}$ && 0.38 $\pm$ 0.29  & 0.43 $\pm$ 0.29  & & 0.42 $\pm$ 0.28  &  0.45 $\pm$ 0.29 \\
                          & Tile no.~(FOV = $1\,\text{deg}^2$)  & 52 $^{+97}_{-51}$ & 47 $^{+77}_{-46}$ && 17 $^{+53}_{-16}$ & 21 $^{+52}_{-20}$ & & 3 $^{+3}_{-2}$ & 4 $\pm$ 3 \\
                          & Tile no.~(FOV = $20\,\text{deg}^2$) & 8 $^{+9}_{-7}$ & 8 $^{+8}_{-7}$ && 2 $^{+7}_{-1}$ & 3 $^{+6}_{-2}$ & & 1 $^{+1}_{-0}$ & 1 $^{+1}_{-0}$ \\
        \bottomrule
    \end{tabular}
    \caption{Tiling and contour level results for BBH, NSBH and BNS type signals, for the scattering glitch injection study.
    $\text{SNR}_{Net}=26$ corresponds to results from injections made only in Livingson and Hanford, \textit{i.e.}\,no Virgo, while $\text{SNR}_{Net}=27$ and $\text{SNR}_{Net}=41$ refer to results with injections made in all three interferometers.
    Tiling results report the number of tile pointings that a telescope would need to search over until the true sky location of an event is observed.
    Contour level corresponds to the skymap probability contour that contains the true GW event location.
    The `Pre-glitch' column refers to injections that were made before the scattering glitch start time $t_0$ (this represents an injection sample that is not affected by the glitch).
    The `Glitch' column refers to injections that were made after the scattering glitch start time $t_0$ (this represents an injection sample that is affected by the glitch).
    The table reports averaged results from the `Pre-glitch' and `Glitch' samples with $1\sigma$ deviation.
    For BBH type signals these are averaged results from $10$ glitches with $20\%$ trimming, while NSBH and BNS type results are from a single representative glitch.
    This also explains why BBH results have generally smaller error bars than NSBH and BNS results.
    }
    \label{tab:sc}
\end{table*}

\subsubsection{BBH}
\label{ssub:sc_bbh}

Even though scattering glitches can differ from each other, we are still able to effectively average BBH results from multiple scattering glitches.
This is due to the fact that our tested BBH-type signals have nearly identical time-frequency content and are very compact (Table~\ref{tab:signal_properties}).

All BBH signals were injected in $4$\,s `Pre-glitch' and `Glitch' windows with the injection spacing of $200$\,ms.
Results reported in Table~\ref{tab:sc} are averaged from 10 glitch runs with $20\%$\ trimming.

Looking at the overall BBH results we can observe that `Pre-glitch' and `Glitch' results are nearly identical except that in some cases the localisation in the presence of a glitch is \textit{better} than in the abscence of a glitch.
We argue that these differences are small enough to be caused by random noise fluctuations.

Out of three BBH events we tested, the biggest difference between `Pre-glitch' and `Glitch' values is for the GW170608-like event.
As an example, $1\,\text{deg}^2$ tiling efficiency at $\text{SNR}_{Net}=27$ is worse by 3 tiles (tile \#8 vs tile \#5).
Further investigation revealed that there is one particular time at the `Pre-glitch' window which skewed results for multiple glitches.
We found no specific reason why sky localisation of a GW170608-like event should be worse at this particular time.

\subsubsection{NSBH}
\label{ssub:sc_nsbh}

We performed injections  for a GW190814-like signal with 10 scattering glitches.
However we found that averaging results from 10 glitches for an extended signal like GW190814 is not practical, just as with thunderstorm glitches (\S\ref{ssub:ts_nsbh}).

Instead we performed one representative run with a scattering glitch that reaches to $70$\,Hz in order to simulate the worst-case scenario.
For this test, both `Pre-glitch' and `Glitch' windows were $3$\,s long with the injection spacing of $50$\,ms.

Table~\ref{tab:sc} shows that the NSBH localisation with small FOV telescopes is affected across all of our tested SNRs. 
For $\text{SNR}_{Net}=26$, tile \#95 instead of tile \#37 contains the true sky location (\textit{i.e.}\,a factor of 2.6 increase), while having three detectors reduces the difference between `Pre-glitch' and `Glitch' results to about the factor of 2.
Large FOV telescopes localising NSBH sources are impacted only with a two-detector network, where tile \#13 instead of tile \#6 contains the true sky location on average.

Skymap contour levels do not change significantly for NSBH sources in the presence of a scattering glitch.

\subsubsection{BNS}
\label{ssub:sc_bns}

As in the NSBH case, we performed the initial injection study with 10 scattering glitches but we found that averaging results for a GW170817-like event is not practical.
As a substitute a single representative run was performed on the same scattering glitch as for the GW190814-like event.
To account for the fact that GW170817-like signal is longer than the GW190814-like signal, the `Pre-glitch' and `Glitch' windows were extended to $8$\,s with the injection spacing of $100$\,ms.

Table~\ref{tab:sc} shows that BNS localisation with two or three detectors is not significantly impacted by scattering glitches.
Tiling results for both, small and large FOV telescopes, are similar between the `Pre-glitch' and `Glitch' windows.
This is also applies to the contour level results.

\section{Discussion}
\label{discussion}

\textit{\textbf{Blips.}} 
Following the coincidence of an extremely loud blip glitch with GW170817~\cite{gw170817}, the autogating procedure was implemented in low latency searches~\cite{pycbc_live_o3}.
This procedure automatically removes the data containing the glitch thus minimizing its effect on estimation of GW source parameters.
However, the autogating is implemented in a way such that only the very loud blip glitches are removed, hence the majority of blip glitches remain a problem for low latency searches.

In our study we found that the sky localisation of BBH signals detected in low latency is affected by blips only at a very specific time relative to the glitch $t_0$, \textit{i.e.}\ $t_0 + 30$\,ms.
This is not surprising given the fact that blip glitches affect high mass BBH searches because of resemblance to such events~\cite{blips, blips_imbh}. 

We estimate that the localisation of a BBH signal can require up to $850$ ($400$) times more tiles to search over in order to find the true sky location of an event with a telescope of FOV=$1\,\text{deg}^2$ ($20\,\text{deg}^2)$.
This means that smaller telescopes suffer from $t_0+30$\,ms bias more than large telescopes.
Yet in both cases the estimated skymap is biased so much that practically it is very unlikely to observe the true sky location.

It is important to note that such bias affects only those BBH signals that have the opposite phase to a blip glitch at $t_0+30$\,ms. 
We also noticed that in some cases the $90\%$ credible area is impacted, \textit{e.g.}\ GW150914-like event (Fig.~\ref{fig:gw150914_skymaps}), but we did not find such evidence for other BBH signals.

Similarly, NSBH low-latency sky localisation is also affected only at $t_0+25$\,ms relative to the blip glitch.
Up to $106$ ($17$) times more tiles need to be searched over to find the true sky location for a FOV=$1\,\text{deg}^2$ ($20\,\text{deg}^2$) telescope.
We found no evidence for any noticeable changes in the $90\%$ credible area at $t_0+25$\,ms.

In contrast to BBH and NSBH signals, the localisation of BNS signals appears to not be impacted by blips at any time.
This happens because longer waveforms, such as GW170817, do not match the phase and amplitude of a blip as well as shorter waveforms of BBH and NSBH signals, thus no destructive interference can occur.

In summary, blip glitches affect low latency sky localisation of NSBH- and BBH-type signals at a very specific time, $t_0+30$\,ms, and only if GW signal overlaps and cancels the blip glitch.

\textit{\textbf{Thunderstorm glitches.}}
GW190521-like events are affected the most from all BBH signals.
For such an event detected by two interferometers, small FOV telescopes may require to observe more than two times of tiles in order to find the true sky location.
Adding a third interferometer allows to reduce the difference in tiling efficiency to only about $50\%$ for small FOV telescopes.
For large FOV telescopes, the localisation of GW190521-like events is impacted only if an event is observed with two interferometers.

Lower mass BBH events, such as GW150914-like and GW170608-like, appear to be impacted by thunderstorm glitches in the abscence of a third interferometer.
In such case both, small and large FOV telescopes are affected.

NSBH signals are significantly affected by thunderstorm glitches.
For events observed with two interferometers there is almost fourfold increase in the number of tiles that need to be searched over using a FOV=$1\,\text{deg}^2$ telescope.
The impact becomes smaller if the event is detected by three interferometers and/or using a larger FOV telescope.

The significance of thunderstorm glitches affecting NSBH localisation is also reflected in the changes to contour level.
In two and three interferometer configurations the contour level increased twofold and sometimes exceeded the 50th percentile, indicating that $50\%$ probability skymaps might not be sufficient localising an NSBH source in the presence of a thunderstorm glitch.

\begin{figure}[htpb]
    \centering
    \includegraphics[width=0.4\paperwidth]{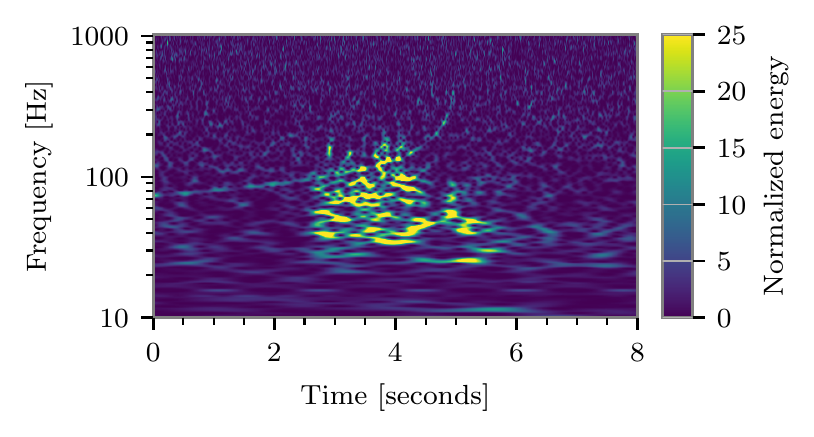}
    \caption{Time-frequency representation of a GW170817-like signal overlapping a thunderstorm glitch. Only a minor part (about $1.7$\,s in total) of the BNS signal coincides with a thunderstorm glitch.}\label{fig:bns_ts_specgram}
\end{figure}

BNS signals can overlap a thunderstorm glitch for multiple seconds (Fig.~\ref{fig:bns_ts_specgram}), yet we find no evidence of any significant change in contour level or tiling results for small and large FOV telescopes (Table~\ref{tab:ts}). 
We argue that this is due to the following reasons:
\begin{enumerate}
    \item A BNS merges at a relatively high frequency ($f_{merger}=4650$\,Hz, Table~\ref{tab:signal_properties}) which means that most of the SNR and sky localisation information is contained at the very late stages of an inspiral, and at very high frequencies\footnote{As an example, only about $20\%$ of sky localisation uncertainty is accumulated $1$\,s before the BNS merger~\cite{bns_loc}.}.
    \item Only a minor part of the whole BNS signal overlaps a glitch. For example Fig.~\ref{fig:bns_ts_specgram} shows about $1.7$\,s overlap of a BNS signal with a thunderstorm glitch.
    \item As we have seen in \S\ref{ssub:blips_bbh}, in order to have a significant effect on sky localisation a glitch needs to overlap and cancel the GW signal. This becomes increasingly difficult with longer duration signals like BNS ($128$\,s for a signal starting at $20$\,Hz) and relatively long glitches like thunderstorms ($3\text{-}10$\,s).
\end{enumerate}

In summary, thunderstorm glitches appear to mostly affect the events detected with two interferometers. 
Observing events with three interferometers and/or using larger FOV telescopes (\textit{e.g.}\,FOV=$20\,\text{deg}^2$) for the EM follow-up campaign reduces the impact of thunderstorm glitches for BBH and NSBH sources.
The localisation of BNS signals seems to not be impacted by thunderstorm glitches.

\textit{\textbf{Scattering glitches}}.
BBH events are not significantly affected by scattering glitches when observed with either two or three interferometers.

The effect of scattering glitches on NSBH signals like GW190814 is noticed only if an event is detected with two interferometers or, in the case of detecting an event with three interferometers, when a relatively small (FOV=$1\,\text{deg}^2$) telescope is used for EM follow-up.

BNS signals like GW170817 appear to not be affected by scattering glitches.
While it is expected that a scattering glitch should overlap a BNS signal, there is no impact on sky localisation, just as with the thunderstorm glitches (\S\ref{ssub:ts_bns}). 
We believe that such results can be explained by the same arguments as in the BNS-thunderstorm case (see above).

In summary, scattering glitches appear to affect only NSBH-type signals.

\section{Conclusions}
\label{conclusions}

Gravitational-wave detectors regularly suffer from non-Gaussian noise artifacts called glitches.
Results from the LIGO-Virgo third observing run suggest that as many as $24\%$ of the observed GW candidate events are in the vicinity of a non-Gaussian noise transient.
This in turn could affect parameter estimation of a GW event, including the sky localisation, whose accuracy is crucial for electromagnetic follow-up efforts.

In this paper we investigated whether sky localisation of a GW event is affected by a nearby glitch in low latency analyses.
Our study examined three classes of glitches: blips, thunderstorms and fast scatterings, and what impact they have on BBH, NSBH and BNS type of GW signals.
Most importantly, our results show that the relative positioning of a glitch with respect to a GW signal can have a drastic impact on GW source sky localisation (albeit in rare cases).
This in turn allows to quantitatively assess the impact of data quality before sending a low latency alert to astronomers, a task that has never been done before.

We found that blip glitches generally do not affect the localisation of BBH, NSBH or BNS signals.
Only in very specific circumstances (\textit{i.e.}\ with the blip glitch positioning of $t_0+30$\,ms for a BBH signal and $t_0+25$\,ms for a NSBH signal) the localisation becomes unreliable.
In such extreme cases the true position of BBH or NSBH may lie well outside the 90\% computed sky-localisation, severely compromising electromagnetic follow-up.
Fortunately, this requires very specific circumstances, thus a low-latency data quality flag could be implemented.
We found no such impact of blip glitches for BNS localisation in low latency.

Thunderstorm glitch results indicate that with two interferometers BBH and NSBH localisation is impacted even using as large as a FOV=$20\,\text{deg}^2$ telescope.
However, if a NSBH or a BBH detection is made with three interferometers, the bias in sky localisation caused by thunderstorm glitches becomes small enough to affect only small (FOV=$1\,\text{deg}^2$) telescopes.
Similarly to blips, our results indicate that thunderstorm glitches do not usually impact BNS event localisation.

Our study suggests that fast scattering glitches have the least impact of the three types of glitches we tested.
Fast scattering glitches have no impact on the low latency localisation of BBH and BNS signals.
For NSBH signals observed with two-detector network, the sky localisation bias due to fast scattering glitches is significant enough to affect even large (FOV=$20\,\text{deg}^2$) telescopes.
Observing NSBH signals with three interferometers reduces the bias such that it impacts only small (FOV=$1\,\text{deg}^2$) telescopes.

Since there is nothing intrinsically specific about blip, thunderstorm and fast scattering glitches, our reported results should be applicable to other glitches that have similar time-frequency content.
As an example, we believe that our blip findings should be applicable to tomte glitches.
At the same time our results should be interpreted carefully and keeping in mind the possible limitations and challenges.
A new observing run could produce a previously unseen glitch type, \textit{e.g.}\,a high frequency long duration glitch, in which case our results should not be applied. 
Further studies could investigate if there is any bias with other types of glitches or even multiple glitches overlapping a GW signal, as well as explore other signal properties important for EM follow-up such as source mass classification.

The fourth LIGO-Virgo-KAGRA observing run is planned to start in December 2022~\cite{kagra_o4}.
In a likely scenario of the glitch rate staying the same (or increasing), we should expect even more GW events to be in the vicinity of non-Gaussian noise transients than during O3 due to increased detector sensitivity.
As our study has shown, glitches can impact the low latency sky localisation of compact binary mergers.
In some cases, \textit{e.g.}~BBH and NSBH signals nearby a short or medium duration glitch, the localisation can be biased and thus a low latency alert should be sent with caution.
At the same time we show that the low latency localisation for BNS signals appears to be not affected by glitches like blips, thunderstorms or fast scatterings, thus an alert can be sent immediately.

\acknowledgments
We thank Michael W.~Coughlin, Tito Dal Canton, Miriam C.~Mueller, Leo P.~Singer, Jonathan Thompson, Tom Dent and members of the LIGO Detector Characterization group for valuable input during the preparation of this manuscript.
We also thank the anonymous referee for valuable comments which improved the manuscript during review.

RM is supported by the STFC grant ST/T000325/1.
LKN thanks the UKRI Future Leaders Fellowship for support through the grant MR/T01881X/1.
JDL acknowledges support from a UK Research and Innovation Fellowship MR/T020784/1.
The authors are grateful for computational resources provided by the LIGO Laboratory and supported by National Science Foundation Grants PHY-0757058 and PHY-0823459.

This research has made use of data, software and/or web tools obtained from the Gravitational Wave Open Science Center (https://www.gw-openscience.org), a service of LIGO Laboratory, the LIGO Scientific Collaboration and the Virgo Collaboration. 
LIGO is funded by the U.S.~National Science Foundation.
Virgo is funded by the French Centre National de Recherche Scientifique (CNRS), the Italian Istituto Nazionale della Fisica Nucleare (INFN) and the Dutch Nikhef, with contributions by Polish and Hungarian institutes.

Signal injection and recovery were performed using \texttt{PyCBC} \citep{pycbc, pycbc_live_o3} and \texttt{LALInference} \cite{lalinference}. 
Skymaps were created and analysed with \texttt{BAYESTAR} \cite{bayestar}, while \texttt{GOTO-tile} \cite{tiling_1} was used for skymap interpretation.
Various stages of the post-processing also used \texttt{GWpy} \cite{gwpy}, \texttt{Astropy} \cite{astropy}, \texttt{SciPy} \cite{scipy}, \texttt{NumPy} \cite{numpy}, \texttt{IPython} \cite{ipython} and \texttt{Matplotlib} \cite{matplotlib}.

This document has been assigned LIGO Laboratory document number P2100460.

\vspace{5mm}
\appendix
\section{Additional results}
\label{appendix}

Tiling, credible area and localizaton distance results for BBH, NSBH, and BNS type signals, for the thunderstorm glitch injection study (Table~\ref{tab:ts_otherFOVs}) and the scattering glitch injection study (Table~\ref{tab:sc_otherFOVs}).

\begin{table*}
    \renewcommand{\arraystretch}{1.3}
    \begin{tabular}{p{2.3cm}p{3.3cm}p{1.9cm}p{1.9cm}p{0.3cm}p{1.9cm}p{1.9cm}p{0.3cm}p{1.9cm}p{1.9cm}} 
        \toprule
         & & \multicolumn{2}{c@{}}{$\text{SNR}_{Net}=26$ (2 detectors)} & & \multicolumn{2}{c@{}}{$\text{SNR}_{Net}=27$} & & \multicolumn{2}{c@{}}{$\text{SNR}_{Net}=41$} \\
        \cmidrule(l){3-4} \cmidrule(l){6-7} \cmidrule(l){9-10} 
          &  & Pre-glitch & Glitch & & Pre-glitch & Glitch & & Pre-glitch & Glitch \\ 
        \midrule
        GW190521-like & Tile no.~(FOV = $0.25\,\text{deg}^2$)  & 249 $\pm$ 218 & 586 $^{+587}_{-385}$ &&  36 $\pm$ 21  & 56 $\pm$ 34 & & 13 $\pm$ 7 & 20 $\pm$ 10\\
                          & Tile no.~(FOV = $5\,\text{deg}^2$) & 15  $\pm$ 13 & 28 $\pm$ 30 &&3 $\pm$ 1 & 4 $\pm$ 2 &  & 1 $^{+1}_{-0}$ & 2 $\pm$ 1 \\
                          & Tile no.~(FOV = $10\,\text{deg}^2$)  & 11 $\pm$ 9 & 19 $\pm$ 18 && 2 $\pm$ 1 & 3 $\pm$ 1 &  & 1 $\pm$ 0  & 1 $^{+1}_{-0}$ \\
                          & Tile no.~(FOV = $40\,\text{deg}^2$) & 5 $\pm$ 4 & 7 $\pm$ 6 && 2 $\pm$ 1 & 2 $\pm$ 1  & & 1 $\pm$ 0  & 1 $^{+1}_{-0}$ \\
                          & $50\%$ credible area (deg$^2$) & 294 $\pm$ 34 & 257 $\pm$ 39 && 13 $\pm$ 3  & 15 $\pm$ 3 && 5 $\pm$ 0 & 5 $\pm$ 1 \\
                          & $90\%$ credible area (deg$^2$) & 1068 $\pm$ 113 & 1060 $\pm$ 155 && 63 $\pm$ 17 & 79 $\pm$ 32 && 17 $\pm$ 2 & 18 $\pm$ 3 \\
                          & Dist$_{P_{\max}}$ (deg) & 3 $^{+6}_{-3}$ & 13 $^{+23}_{-13}$ && 2 $\pm$ 1 & 2 $\pm$ 1 && 1 $\pm$ 0 & 1 $\pm$ 0 \\
                          \midrule
        GW150914-like & Tile no.~(FOV = $0.25\,\text{deg}^2$)  & 156 $\pm$ 134 & 293  $\pm$ 267  &&  49 $\pm$ 36  & 50 $\pm$ 40 & & 12 $\pm$ 7 & 13 $\pm$ 7\\
                          & Tile no.~(FOV = $5\,\text{deg}^2$)  & 9 $\pm$ 7 & 18 $\pm$ 17 && 4 $\pm$ 2 & 3 $\pm$ 2 &  & 1 $^{+1}_{-0}$ & 1 $^{+1}_{-0}$ \\
                          & Tile no.~(FOV = $10\,\text{deg}^2$)  & 8 $\pm$ 6 & 15 $\pm$ 11 && 3 $\pm$ 2 & 3 $\pm$ 1 &  & 1 $\pm$ 0  & 1 $\pm$ 0 \\
                          & Tile no.~(FOV = $40\,\text{deg}^2$) & 4 $\pm$ 3 & 6 $\pm$ 4 && 2 $\pm$ 1 & 2 $\pm$ 1  & & 1 $\pm$ 0  & 1 $\pm$ 0 \\
                          & $50\%$ credible area (deg$^2$) & 169 $\pm$ 25 & 174 $\pm$ 26 && 19 $\pm$ 8 & 19 $\pm$ 10 && 4 $\pm$ 1 & 3 $\pm$ 1 \\
                          & $90\%$ credible area (deg$^2$) & 717 $\pm$ 64 & 704 $\pm$ 72 && 127 $\pm$ 54 & 130 $\pm$ 66 && 15 $\pm$ 5 & 14 $\pm$ 4 \\
                          & Dist$_{P_{\max}}$ (deg) & 3 $^{+7}_{-3}$ & 13 $^{+23}_{-13}$ && 3 $\pm$ 2 & 3 $\pm$ 3 && 1 $\pm$ 0 & 1 $\pm$ 0 \\
                          \midrule
        GW170608-like & Tile no.~(FOV = $0.25\,\text{deg}^2$)  & 152 $\pm$ 130 & 180 $\pm$ 153 &&  21 $\pm$ 13  & 20 $\pm$ 12 & & 7 $\pm$ 3 & 7 $\pm$ 4\\
                          & Tile no.~(FOV = $5\,\text{deg}^2$)  & 11 $\pm$ 10 & 12 $\pm$ 10 && 2 $\pm$ 1 & 2 $\pm$ 1 &  & 1 $\pm$ 0  & 1 $\pm$ 0 \\
                          & Tile no.~(FOV = $10\,\text{deg}^2$)  & 9 $\pm$ 6 & 10 $\pm$ 7 && 2 $\pm$ 1 & 2 $\pm$ 1 &  & 1 $\pm$ 0  & 1 $\pm$ 0 \\
                          & Tile no.~(FOV = $40\,\text{deg}^2$) & 5 $\pm$ 3 & 5 $\pm$ 3 && 2 $\pm$ 1 & 2 $\pm$ 1  & & 1 $\pm$ 0  & 1 $\pm$ 0 \\
                          & $50\%$ credible area (deg$^2$) & 77 $\pm$ 12 & 77 $\pm$ 13 && 5 $\pm$ 1 & 6 $\pm$ 2 && 2 $\pm$ 0 &  2 $\pm$ 0\\
                          & $90\%$ credible area (deg$^2$) & 319 $\pm$ 42 & 313 $\pm$ 48 && 27 $\pm$ 10 & 28 $\pm$ 11 && 6 $\pm$ 1 & 6 $\pm$ 1 \\
                          & Dist$_{P_{\max}}$ (deg) & 5 $^{+7}_{-5}$ & 7 $^{+11}_{-7}$ && 1 $\pm$ 1 & 1 $\pm$ 1 && 1 $\pm$ 0 & 1 $\pm$ 0 \\
                          \midrule
        GW190814-like & Tile no.~(FOV = $0.25\,\text{deg}^2$)  & 312 $^{+669}_{-311}$ &  1106 $^{+1501}_{-1105}$ &&  70$^{224}_{-69}$  & 160 $^{+432}_{-159}$ & & 10 $^{+10}_{-9}$ & 19 $\pm^{+29}_{-18}$\\
                          & Tile no.~(FOV = $5\,\text{deg}^2$)  & 16 $^{+29}_{-15}$ & 59 $^{+82}_{-58}$ && 4 $^{+12}_{-3}$ & 9 $^{+22}_{-8}$ &  & 1 $\pm$ 0  & 1 $^{+1}_{-0}$ \\
                          & Tile no.~(FOV = $10\,\text{deg}^2$)  & 11 $^{+18}_{-10}$ & 34 $^{+45}_{-33}$ && 3 $^{+6}_{-2}$ & 5 $^{+12}_{-4}$ &  & 1 $\pm$ 0  & 1 $^{+1}_{-0}$ \\
                          & Tile no.~(FOV = $40\,\text{deg}^2$) & 4 $^{+5}_{-3}$ & 12 $^{+18}_{-11}$ && 2 $^{+3}_{-1}$ & 3 $^{+5}_{-2}$  & & 1 $\pm$ 0  & 1 $^{+1}_{-0}$ \\
                          & $50\%$ credible area (deg$^2$) & 115 $\pm$ 32 & 117 $\pm$ 31  && 12$^{+15}_{-12}$ & 16 $^{+17}_{-16}$ && 3 $\pm$ 1 &  3 $\pm$ 2\\
                          & $90\%$ credible area (deg$^2$) & 473 $\pm$ 107 & 465 $\pm$ 114 && 64$^{+70}_{-64}$ & 80 $\pm$ 80 && 10 $\pm$ 3 & 13 $\pm$ 11 \\
                          & Dist$_{P_{\max}}$ (deg) & 22 $^{+48}_{-22}$ & 45 $^{+59}_{-45}$ && 7 $^{+27}_{-7}$ & 17 $^{+45}_{-17}$ && 1 $\pm$ 1 & 1 $\pm$ 1 \\
                          \midrule
        GW170817-like & Tile no.~(FOV = $0.25\,\text{deg}^2$)  & 217 $^{+392}_{-216}$ & 218 $^{+345}_{-217}$ &&  107 $^{+278}_{-106}$  & 95 $\pm^{+254}_{-94}$ & & 9 $\pm^{+9}_{-9}$ & 8 $\pm^{+10}_{-8}$ \\
                          & Tile no.~(FOV = $5\,\text{deg}^2$)  & 11 $^{+19}_{-10}$ & 11 $^{+18}_{-10}$ && 7 $^{+18}_{-6}$ & 6 $^{+16}_{-5}$ &  & 1 $^{+1}_{-0}$  & 1 $^{+1}_{-0}$  \\
                          & Tile no.~(FOV = $10\,\text{deg}^2$)  & 9 $^{+14}_{-8}$ & 9 $^{+12}_{-8}$ && 5 $^{+12}_{-4}$ & 4 $^{+11}_{-3}$ &  & 1 $^{+1}_{-0}$ & 1 $^{+1}_{-0}$  \\
                          & Tile no.~(FOV = $40\,\text{deg}^2$) & 3 $^{+5}_{-2}$ & 3 $^{+4}_{-2}$ && 3 $^{+5}_{-2}$ & 3 $^{+5}_{-2}$  & & 1 $^{+1}_{-0}$  & 1 $^{+1}_{-0}$  \\
                          & $50\%$ credible area (deg$^2$) & 76 $\pm$ 25 & 81 $\pm$ 26  && 8 $^{+9}_{-8}$ &  8 $^{+10}_{-8}$ && 2 $\pm$ 1 &  2 $\pm$ 2 \\
                          & $90\%$ credible area (deg$^2$) & 312 $\pm$ 74 & 313 $\pm$ 75 && 45$^{+47}_{-45}$ & 46 $^{+49}_{-46}$ && 8 $^{+9}_{-8}$ & 9 $^{+15}_{-9}$ \\
                          & Dist$_{P_{\max}}$ (deg) & 12 $^{+36}_{-12}$ & 31 $^{+58}_{-31}$ && 14 $^{+39}_{-14}$ & 15 $^{+42}_{-15}$ && 1 $^{+2}_{-1}$ & 1 $^{+5}_{-1}$ \\
         \bottomrule
    \end{tabular}
    \caption{Tiling, credible area and localisaton distance results for BBH, NSBH and BNS type signals, for the thunderstorm glitch injection study.
    $\text{SNR}_{Net}=26$ corresponds to results from injections made only in Livingson and Hanford, \textit{i.e.}\,no Virgo, while $\text{SNR}_{Net}=27$ and $\text{SNR}_{Net}=41$ refer to results with injections made in all three interferometers.
    Tiling results report the number of tile pointings that a telescope would need to search over until the true sky location of an event is observed.
    Localisation distance Dist$_{P_{\max}}$ is the angular distance between the true sky location of a GW event and the maximum probability pixel in the corresponding skymap.
    The `Pre-glitch' column refers to injections that were made before the thunderstorm glitch start time $t_0$ (this represents an injection sample that is not affected by the glitch).
    The `Glitch' column refers to injections that were made after the thunderstorm glitch start time $t_0$ (this represents an injection sample that is affected by the glitch).
    The table reports averaged results from the `Pre-glitch' and `Glitch' samples with $1\sigma$ deviation.
    For BBH type signals these are averaged results from $10$ glitches with $20\%$ trimming, while NSBH and BNS type results are from a single representative glitch.
    This also explains why BBH results have generally smaller error bars than NSBH and BNS results.
    }
    \label{tab:ts_otherFOVs}
\end{table*}

\begin{table*}
    \renewcommand{\arraystretch}{1.3}
    \begin{tabular}{p{2.3cm}p{3.3cm}p{1.9cm}p{1.9cm}p{0.3cm}p{1.9cm}p{1.9cm}p{0.3cm}p{1.9cm}p{1.9cm}} 
        \toprule
         & & \multicolumn{2}{c@{}}{$\text{SNR}_{Net}=26$ (2 detectors)} & & \multicolumn{2}{c@{}}{$\text{SNR}_{Net}=27$} & & \multicolumn{2}{c@{}}{$\text{SNR}_{Net}=41$} \\
        \cmidrule(l){3-4} \cmidrule(l){6-7} \cmidrule(l){9-10} 
          &  & Pre-glitch & Glitch & & Pre-glitch & Glitch & & Pre-glitch & Glitch \\ 
        \midrule
        GW190521-like & Tile no.~(FOV = $0.25\,\text{deg}^2$)  & 239 $\pm$ 223 & 250 $^{+253}_{-249}$ &&  55 $\pm$ 35  & 48 $\pm$ 27 & & 20 $\pm$ 11 & 16 $\pm$ 7 \\
                          & Tile no.~(FOV = $5\,\text{deg}^2$)  & 12 $\pm$ 10 & 14 $\pm$ 13 && 3 $\pm$ 2 & 3 $\pm$ 2 &  & 2 $\pm$ 1  & 2 $\pm$ 1 \\
                          & Tile no.~(FOV = $10\,\text{deg}^2$)  & 8 $^{+8}_{-7}$ & 9 $\pm$ 8 && 2 $\pm$ 1 & 2 $\pm$ 1 &  & 1 $\pm$ 1  & 1 $\pm$ 1 \\
                          & Tile no.~(FOV = $40\,\text{deg}^2$) & 6 $\pm$ 4 & 6 $\pm$ 5 && 2 $\pm$ 1 & 2 $\pm$ 1  & & 1 $\pm$ 0  & 1 $\pm$ 0 \\
                          & $50\%$ credible area (deg$^2$) & 257 $\pm$ 35 & 268 $\pm$ 39 && 16 $\pm$ 3 & 15 $\pm$ 3 && 5 $\pm$ 1 & 5 $\pm$ 1 \\
                          & $90\%$ credible area (deg$^2$) & 1131 $\pm$ 138 & 1118 $\pm$ 129 && 69 $\pm$ 15 & 73 $\pm$ 24 && 19 $\pm$ 4 & 18 $\pm$ 4 \\
                          & Dist$_{P_{\max}}$ (deg) & 1 $\pm$ 0 & 3 $^{+6}_{-3}$ && 2 $\pm$ 1 & 2 $\pm$ 2 && 1 $\pm$ 0 & 1 $\pm$ 0 \\
                          \midrule
        GW150914-like & Tile no.~(FOV = $0.25\,\text{deg}^2$)  & 119 $^{+128}_{-118}$ & 116 $^{+129}_{-115}$ &&  24 $\pm$ 18  & 16 $\pm$ 9 & & 8 $\pm$ 5 & 6 $\pm$ 3 \\
                          & Tile no.~(FOV = $5\,\text{deg}^2$)  & 8 $\pm$ 7 & 8 $\pm$ 7 && 2 $\pm$ 1 & 2 $\pm$ 1 &  & 1 $\pm$ 0  & 1 $\pm$ 0 \\
                          & Tile no.~(FOV = $10\,\text{deg}^2$)  & 5 $^{+5}_{-4}$ & 6 $\pm$ 5 && 2 $\pm$ 1 & 1 $\pm$ 1 &  & 1 $\pm$ 0  & 1 $\pm$ 0 \\
                          & Tile no.~(FOV = $40\,\text{deg}^2$) & 5 $\pm$ 4 & 5 $\pm$ 3 && 2 $\pm$ 1 & 1 $\pm$ 0  & & 1 $\pm$ 0  & 1 $\pm$ 0 \\
                          & $50\%$ credible area (deg$^2$) & 114 $\pm$ 15 & 118 $\pm$ 16  && 6 $\pm$ 2 & 6 $\pm$ 2 && 2 $\pm$ 0 & 2 $\pm$ 0 \\
                          & $90\%$ credible area (deg$^2$) & 483 $\pm$ 62 & 509 $\pm$ 60  && 31 $\pm$ 16 & 37 $\pm$ 23 && 7 $\pm$ 1 & 7 $\pm$ 1 \\
                          & Dist$_{P_{\max}}$ (deg) & 1 $\pm$ 0 & 2 $^{+4}_{-2}$ && 1 $\pm$ 1 & 1 $\pm$ 0 && 1 $\pm$ 0 & 1 $\pm$ 0 \\
                          \midrule
        GW170608-like & Tile no.~(FOV = $0.25\,\text{deg}^2$)  & 233 $\pm$ 186 & 192 $\pm$ 165 &&  31 $\pm$ 24  & 20 $\pm$ 12 & & 10 $\pm$ 6 & 9 $\pm$ 4 \\
                          & Tile no.~(FOV = $5\,\text{deg}^2$)  & 11 $\pm$ 10 & 9 $^{+9}_{-8}$ && 2 $\pm$ 1 & 2 $\pm$ 1 &  & 1 $\pm$ 0  & 1 $\pm$ 0 \\
                          & Tile no.~(FOV = $10\,\text{deg}^2$)  & 7 $\pm$ 5 & 7 $\pm$ 5 && 2 $\pm$ 1 & 2 $\pm$ 1 &  & 1 $\pm$ 0  & 1 $\pm$ 0 \\
                          & Tile no.~(FOV = $40\,\text{deg}^2$) &5 $\pm$ 3  & 5 $\pm$ 3 && 2 $\pm$ 1 & 1 $\pm$ 0  & & 1 $\pm$ 0  & 1 $\pm$ 0 \\
                          & $50\%$ credible area (deg$^2$) & 80 $\pm$ 13 & 82 $\pm$ 12 && 6 $\pm$ 2 & 6 $\pm$ 2 && 2 $\pm$ 0 &  2 $\pm$ 0 \\
                          & $90\%$ credible area (deg$^2$) & 334 $\pm$ 47 & 338 $\pm$ 42 && 32 $\pm$ 14 & 29 $\pm$ 12 && 6 $\pm$ 1 & 6 $\pm$ 1 \\
                          & Dist$_{P_{\max}}$ (deg) & 2 $^{+4}_{-2}$ & 6 $^{+9}_{-6}$ && 1 $\pm$ 1 & 1 $\pm$ 1 && 1 $\pm$ 0 & 1 $\pm$ 0 \\
                          \midrule
        GW190814-like & Tile no.~(FOV = $0.25\,\text{deg}^2$)  & 192 $^{+325}_{-191}$ & 413 $^{+1056}_{-412}$ &&  42 $\pm^{+81}_{-42}$  & 68 $\pm^{+136}_{-68}$ & & 11 $\pm$ 11 & 26 $\pm^{+98}_{-26}$ \\
                          & Tile no.~(FOV = $5\,\text{deg}^2$)  & 7 $^{+10}_{-6}$ & 19 $^{+48}_{-18}$ && 3 $\pm^{+4}_{-3}$ & 4 $\pm^{+6}_{4}$ &  & 1 $\pm$ 1  & 2 $\pm^{+7}_{-2}$ \\
                          & Tile no.~(FOV = $10\,\text{deg}^2$)  & 4 $^{+5}_{-3}$ & 11 $^{+26}_{-10}$ && 2 $\pm^{+2}_{-2}$ & 2 $\pm^{+4}_{-2}$ &  & 1 $\pm$ 0  & 2 $\pm^{+4}_{-2}$ \\
                          & Tile no.~(FOV = $40\,\text{deg}^2$) & 4 $\pm$ 3 & 8 $^{+12}_{-7}$ && 2 $\pm$ 1 & 2 $\pm^{+2}_{-2}$  & & 1 $\pm$ 0  & 1 $\pm^{+2}_{-1}$ \\
                          & $50\%$ credible area (deg$^2$) & 101 $\pm$ 29 &124 $\pm$ 43  &&  & 19 $\pm$ 48 $^{+14}_{-12}$ && 3 $\pm$ 1 & 3 $\pm$ 2 \\
                          & $90\%$ credible area (deg$^2$) & 426 $\pm$ 94 & 498 $\pm$ 130 && 55$^{+63}_{-55}$ & 69 $^{+73}_{-69}$ && 9 $\pm$ 4 & 11 $\pm$ 9 \\
                          & Dist$_{P_{\max}}$ (deg) & 10 $^{+33}_{-10}$ & 13 $^{+33}_{-13}$ && 7 $^{+27}_{-7}$ & 12 $^{+37}_{-12}$ && 1 $\pm$ 1 & 3 $^{+19}_{-3}$ \\
                          \midrule
        GW170817-like & Tile no.~(FOV = $0.25\,\text{deg}^2$)  & 278 $^{+519}_{-277}$ & 242 $^{+402}_{-241}$  &&  67 $\pm^{+198}_{-67}$  & 83 $\pm^{+219}_{-83}$ & & 11 $\pm^{+13}_{-11}$ & 13 $\pm^{+15}_{-13}$ \\
                          & Tile no.~(FOV = $5\,\text{deg}^2$)  & 10 $^{+19}_{-9}$ & 9 $^{+14}_{-8}$ && 4 $\pm^{+13}_{-4}$ & 5.0 $\pm^{+12}_{-5}$ &  & 1 $\pm$ 1  & 1 $\pm$ 1 \\
                          & Tile no.~(FOV = $10\,\text{deg}^2$)  & 6 $^{+10}_{-5}$ & 5 $^{+8}_{-4}$ && 3 $\pm^{+7}_{-3}$ & 3.3 $\pm^{+8}_{-3}$ &  & 1 $\pm$ 0  & 1 $\pm$ 0 \\
                          & Tile no.~(FOV = $40\,\text{deg}^2$) & 5 $^{+5}_{-4}$ & 5 $^{+5}_{-4}$ && 2 $\pm^{+4}_{-2}$ & 2.6 $\pm^{+4}_{-3}$  & & 1 $\pm$ 0  & 1 $\pm$ 0 \\
                          & $50\%$ credible area (deg$^2$) & 73 $\pm$ 22 & 80 $\pm$ 22 && 2 $\pm$ 1 & 2 $\pm$ 1 &&  7 $\pm$ 7 & 8 $\pm$ 8 \\
                          & $90\%$ credible area (deg$^2$) & 304 $\pm$ 76 & 332 $\pm$ 75 && 38 $\pm$ 37 & 43 $^{+45}_{-43}$ && 7$^{+9}_{-7}$ & 7 $\pm$ 7 \\
                          & Dist$_{P_{\max}}$ (deg) & 18 $^{+44}_{-18}$ & 18 $^{+45}_{-18}$ && 1 $\pm$ 1 & 1 $^{+2}_{-1}$ && 6 $^{+24}_{-6}$ & 15 $^{+41}_{-15}$ \\
         \bottomrule
    \end{tabular}
    \caption{Tiling, credible area and localisaton distance results for BBH, NSBH and BNS type signals, for the scattering glitch injection study.
    $\text{SNR}_{Net}=26$ corresponds to results from injections made only in Livingson and Hanford, \textit{i.e.}\,no Virgo, while $\text{SNR}_{Net}=27$ and $\text{SNR}_{Net}=41$ refer to results with injections made in all three interferometers.
    Tiling results report the number of tile pointings that a telescope would need to search over until the true sky location of an event is observed.
    Localisation distance Dist$_{P_{\max}}$ is the angular distance between the true sky location of a GW event and the maximum probability pixel in the corresponding skymap.
    The `Pre-glitch' column refers to injections that were made before the scattering glitch start time $t_0$ (this represents an injection sample that is not affected by the glitch).
    The `Glitch' column refers to injections that were made after the scattering glitch start time $t_0$ (this represents an injection sample that is affected by the glitch).
    The table reports averaged results from the `Pre-glitch' and `Glitch' samples with $1\sigma$ deviation.
    For BBH type signals these are averaged results from $10$ glitches with $20\%$ trimming, while NSBH and BNS type results are from a single representative glitch.
    This also explains why BBH results have generally smaller error bars than NSBH and BNS results.
    }
    \label{tab:sc_otherFOVs}
\end{table*}

\end{document}